\documentclass[sigplan,10pt]{acmart}
\renewcommand\footnotetextcopyrightpermission[1]{}
\usepackage{algorithm}
\usepackage{algorithmicx}
\usepackage{algpseudocode}
\algdef{SE}[DOWHILE]{Do}{doWhile}{\algorithmicdo}[1]{\algorithmicwhile\ #1}%
\usepackage{breqn}
\usepackage{makecell}
\usepackage[normalem]{ulem}
\usepackage{pifont}
\usepackage{subfigure}
\usepackage{cleveref}
\useunder{\uline}{\ul}{}
\AtBeginDocument{%
	}

\setcopyright{acmlicensed}
\copyrightyear{2018}
\acmYear{2018}
\acmDOI{XXXXXXX.XXXXXXX}

\acmConference[Conference acronym 'XX]{Make sure to enter the correct
	conference title from your rights confirmation emai}{June 03--05,
	2018}{Woodstock, NY}
\acmISBN{978-1-4503-XXXX-X/18/06}

\newcommand{\pname}[1]{Lynx{#1}}
\newcommand{\heu}[1]{Lynx-HEU{#1}}
\newcommand{\opt}[1]{Lynx-OPT{#1}}

\newcommand{\wj}[1]{{\color{black}#1}}

\newcommand{\fast}[1]{{\color{black}#1}}
\newcommand{\osdi}[1]{{\color{black}#1}}

\begin{document}
	
	\title{Optimizing Large Model Training through Overlapped Activation Recomputation}
	\newcommand{\cp}[1]{{\color{black} #1}}
	\author{{Ping Chen, Wenjie Zhang, Shuibing He, Weijian Chen, Siling Yang, Kexin Huang, Yanlong Yin, Xuan Zhan, Yingjie Gu$^{\dagger}$, Zhuwei Peng$^{\dagger}$, Yi Zheng$^{\dagger}$, Zhefeng Wang$^{\dagger}$, Gang Chen}\\
		{Zhejiang University,  Huawei Cloud$^{\dagger}$}\\
		}

	\begin{abstract}

	Large model training often uses recomputation to alleviate memory
pressure and pipelines to exploit the parallelism of data, tensors, and devices. 
However, existing recomputation approaches may incur high overhead
when training real-world models,
as they are executed on demand in the critical training path.
In this paper, we present \pname{}, a new recomputation framework to
reduce overhead by overlapping recomputation with communication
in training pipelines. 
To reduce the large search space for recomputation strategies,
we propose a heuristic-based recomputation scheduling algorithm,
which is based on the observation
that there are identical structures in large DNN models so that we can apply
the same scheduling policy to all such structures.
Additionally, we propose a recomputation-aware model partitioning method to balance each stage's execution time for improved training throughput.
Our comprehensive evaluation using
GPT models with 1.3B--23B parameters shows that \pname{} outperforms 
existing recomputation approaches by up to 1.37$\times$.

	\end{abstract}


	\settopmatter{printfolios=true}
	\maketitle
	\pagestyle{plain}

\section{Introduction}

\textbf{Motivation.}
Deep neural networks (DNNs) have excelled in domains like natural language processing~\cite{attention}, computer vision~\cite{gptimage-icml20}, and text-to-video generation~\cite{videogpt}. 
Scaling laws~\cite{scaling} reveal that larger models achieve better performance, driving a rapid increase in model sizes.
For instance, from GPT-2 (1.5B parameters, 2019~\cite{gpt2}) to PaLM (540B parameters, 2022~\cite{palm}), model sizes have grown over 360$\times$. This trend is expected to continue~\cite{scaling}, with the growth far outpacing the memory capacity of individual GPUs, typically limited to tens of GBs.

Training such massive models requires parallelization across multiple GPUs~\cite{MegaScale-arxiv24,opt-arxiv22} using techniques like pipeline parallelism~\cite{Gpipe-NIPS19,Pipedream} and tensor parallelism~\cite{Megatron-Arxiv19}. However, even these methods struggle with GPU memory limits. For example, attempting to train GPT 7B with a batch size of 32 on eight NVIDIA A100 GPUs (each with 40GB of memory) results in out-of-memory failures despite employing both tensor and pipeline parallelism. This highlights the urgent need for more efficient memory management techniques.

To address GPU memory limitations, recomputation techniques have emerged as a promising solution.
By discarding activations generated during forward propagation and regenerating them during backpropagation, these methods reduce memory consumption significantly~\cite{checkpointing-arxiv16}.
Recomputation is now widely adopted in frameworks like Megatron-LM~\cite{megatron-ckpt}, MindSpore~\cite{MindSpore}, and Colossal-AI~\cite{colossalai}, each using specific policies to decide which tensors to retain and which to recompute.

\textbf{Limitations of existing recomputation approaches.}
The existing recomputation methods can be placed into two categories and introduce several problems.
(1) \emph{Rule-based recomputation methods~\cite{megatron-ckpt} lack adaptivity.}
These methods rely on predefined patterns that ignore available GPU memory and model-specific requirements.
This lack of adaptivity often leads to excessive recomputation, inefficient memory utilization, and the need for manual tuning (\cref{sec:limitations}).
(2) \emph{Model-adaptive methods like Checkmate~\cite{checkmate19} have poor scalability.}
These approaches use optimization algorithms to tailor recomputation policies to specific models. However, their scalability is limited, as large models create vast search spaces that are computationally expensive to navigate.
(3) \emph{Both rule-based and model-adaptive methods incur significant latency} because all recomputation is performed on the critical training path~\cite{activation-recomputation-MLsys23}.

\begin{figure}
	\centering
	\includegraphics[width=3.2in]{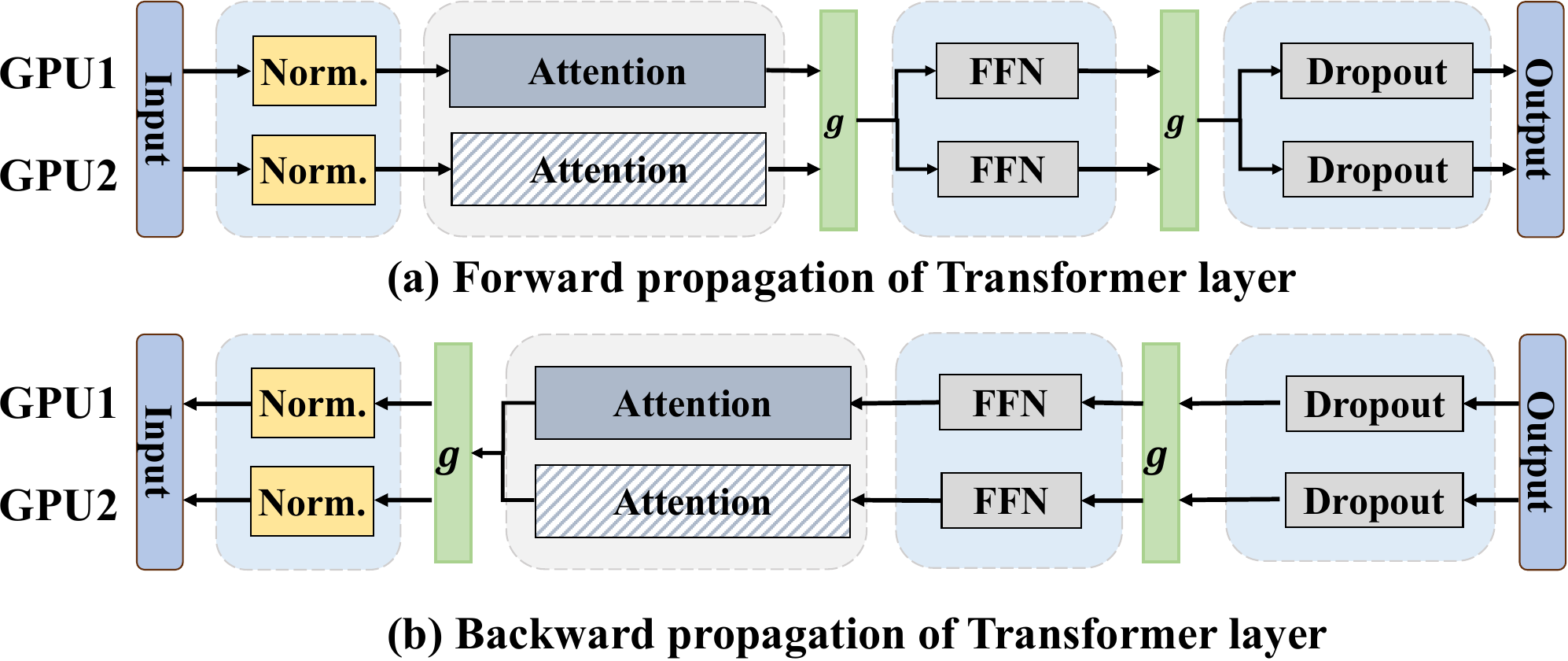}
	\caption{The training workflow of tensor parallelism. The shaded rectangle indicates the splitting of the tensor onto another GPU for parallel training. $g$ denotes the all-reduce operation in the forward and backward.}
	\label{fig:tp}
\end{figure}

\textbf{Observations.}
\osdi{We have three observations in the paper.}
First, tensor parallelism introduces substantial communication overhead, with all-reduce operations (in Figure~\ref{fig:tp}) between GPUs wasting  10\%--70\% of training time (\S\ref{sec:opportunities}). 
Second, in pipeline parallelism, memory usage across different pipeline stages is imbalanced, with earlier stages consuming up to 1.5$\times$ more memory than later stages (\S\ref{sec:opportunities}). 
Finally, recomputation operations can be scheduled flexibly before backward propagation, offering opportunities to optimize their timing and overlap with other operations (Figure~\ref{fig:recomputation}). But existing methods fail to fully exploit this opportunity.

\textbf{Our work.}
The aforementioned three findings motivate us to propose a new recomputation framework. Our design goals are (1) overlapping recomputation with communication to minimize recomputation overhead, (2) optimizing GPU memory utilization by selectively storing tensors in memory to prevent unnecessary recomputation, (3) achieving load balancing across pipeline stages. 
To achieve these goals, we introduce two algorithms to determine recomputation scheduling policy considering which tensor should be recomputed, when they will be recomputed, and how to overlap them with communication.

The first algorithm achieves a
global optimum by searching the whole solution space. We
named it \opt{}. It is modeled as a mixed-integer linear program.
While \opt{} provides an upper bound of training performance,
it cannot be used for scheduling for large models because its
search time is exponentially increased with the model size
(\S\ref{sec:optimal}).

To solve this challenge, we design a heuristic-based recomputation scheduling algorithm (\heu{}) based on the observation that there are identical structures in large DNN models
and local optimal scheduling policy obtained for one layer can
be used for other layers with the same structure  (\S\ref{sec:heuristic}).
\heu{} can be modeled as an integer linear program. Our results show that \heu{} has search time of
seconds and achieves near-optimal performance (\S\ref{eval:heu}). For achieving load balancing among pipeline stages, we design a greedy algorithm for model partitioning  (\S\ref{sec:partition}). None of the existing partitioning algorithms work in our scenario because they do not consider overlapping recomputation with communication
in training pipelines. Our partitioning algorithm iteratively
searches for better results and terminates upon achieving load
balance.

\textbf{Contributions.} In summary, we make the following contributions: (1) To the best of our knowledge, Lynx is the first recomputation framework that fully explores the potential of overlapping recomputation with communication and utilizing idle GPU memory to eliminate unnecessary tensor recomputation. (2) We introduce \pname{}-OPT and \pname{}-HEU for searching recomputation scheduling policy and devise a recomputation-aware model partitioning algorithm to ensure load balancing across pipeline stages, thereby maximizing training throughput. (3) We conduct comprehensive evaluation, demostrating that
\pname{} achieves up to a 1.37$\times$ improvement over existing recomputation methods.

\section{Background and Motivation}

\subsection{Large Model Training}
\label{sec:background}

Deep learning models are built with layers and iteratively trained using batches of samples. 
Each training step consists of forward propagation (FP) and backward propagation (BP), which refine the model's parameters to enhance accuracy.
Activations are intermediate outputs generated during FP and are utilized by BP for gradient calculation.
During FP, input activations, together with the current layer's weights and biases, generate output activations, which serve as the input data for the subsequent layer. 
BP starts from the output layer and traverses layers in reverse to optimize the weights and biases.
To improve throughput and device utilization, training typically processes samples in large batches~\cite{PTD-P-SC21,FlashNeuron-FAST21,icache}.

To accelerate training, large models are parallelized across multiple GPUs.
For example, training GPT-3 (175B parameters) requires 355 GPU-years~\cite{GPT-3}, 
OPT-175B uses 992 80GB A100 GPUs~\cite{opt-arxiv22}, and ByteDance's 175B model employs 12,288 GPUs~\cite{MegaScale-arxiv24}.
To efficiently utilize training devices, data parallelism (DP), tensor parallelism (TP), and pipeline parallelism (PP) have been proposed, and become the state-of-the-art distributed training methods~\cite{MegaScale-arxiv24,Merak-TPDS23,Megatron-Arxiv19,opt-arxiv22}.

\textbf{Data parallelism.}
DP accelerates training by distributing input samples across multiple workers, each of which holds a replica of the model.
By splitting large batches among GPUs, DP enables faster training~\cite{PyTorch,Tensorflow-OSDI16,zico}.

\begin{figure}
	\centering
	\includegraphics[width=3.2in]{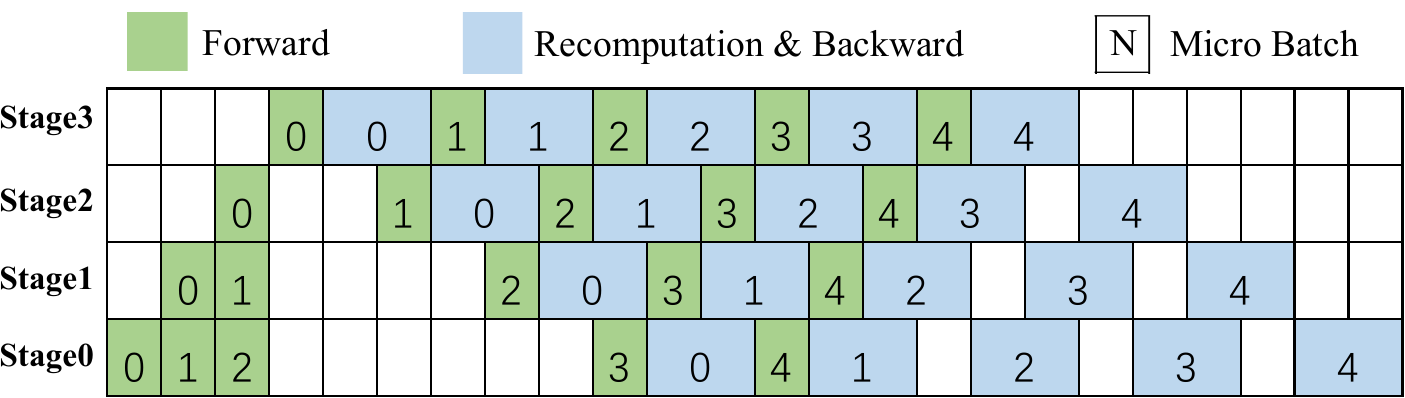}
	\caption{The training workflow of pipeline parallelism (one-forward-one-backward). Each minibatch consists of 5 micro batches. The example illustrates that ideal computation-balanced model partitioning achieves the best training performance.}
	\label{fig:pp}
	\vspace{-0.1in}
\end{figure}

\textbf{Tensor parallelism.}
TP addresses the challenge of accommodating large models by splitting model layers across multiple GPUs~\cite{Megatron-Arxiv19}.
As shown in Figure~\ref{fig:tp}, it parallelizes model parameters, optimizer states inside the attention and MLP blocks, and activations on GPUs. 
During training, it introduces two all-reduce communication operations in both the forward and backward passes to collect the computing result from each GPU to ensure training correctness. 

\textbf{Pipeline parallelism.}
PP splits a model into sub-modules and maps them to multiple GPUs. Outputs from one submodule are transferred to the GPUs hosting the subsequent stage. A batch is split into smaller micro batches, which are processed as a stream in a pipeline, to maximize device utilization.
Given the substantial memory demands during large model training, the mainstream systems often employ a one-forward-one-backward (1F1B) training mechanism~\cite{Pipedream,Pipedram-2-PMLR21,DAPPLE-PPoPP21,activation-recomputation-MLsys23}. In this approach, each pipeline stage alternates between FP and BP for micro-batches. For optimal performance, all pipeline stages should have similar execution times as shown in Figure~\ref{fig:pp}. Otherwise, stalls between stages may occur due to imbalanced load distribution~\cite{MPress-HPCA23}.

\textbf{Impact of GPU memory.}
The limited memory capacity of GPUs imposes significant constraints on large model training.
Specifically, memory is required to
manage both model states and activations (feature maps).
Model states comprise parameters, gradients, and optimizer 
states, such as momentum and variances in Adam~\cite{Adam}. 
A model with $n$ parameters requires 16$n$ bytes of memory, 
including FP16 parameters (2$n$ bytes), one copy of FP16 gradients (2$n$ bytes), 
and FP32 optimizer data (4$n$ bytes each for momentum, variances,
and parameters).
The memory consumption for activations depends on the batch 
size. Users often employ a large batch size to maximize 
GPU utilization~\cite{FlashNeuron-FAST21}, resulting in 
significant memory consumption during training.
For instance, training a 4.7B GPT model on 8 A100 GPUs (TP=8) with 
a batch size of 4 per GPU requires 8GB for model states and 
7.6GB for activations, leading to a GPU utilization~\cite{GPU_Util} 
of 74\%. 
Increasing the batch size to 8 raises GPU utilization to 89\%, but also increases memory usage by 45\% during training.

\subsection{Limitations of Existing Solutions}
\label{sec:limitations}
Activation recomputation (or activation checkpointing) is one of the major 
approaches used for training large models with limited GPU 
memory~\cite{checkpointing-arxiv16,checkmate19,Megatron-Arxiv19,activation-recomputation-MLsys23}.
It discards activation tensors after their final use in the forward pass and then
recomputes them as required during the backward pass.
However, due to the inherent complexity of large model training~\cite{survey-TPDS23}, 
existing efforts have the following weaknesses, which are summarized 
in Table~\ref{table:comp}.

\wj{
\textbf{1. Rule-based methods have poor adaptivity.}
\textit{Full Recomputation} method used in the mainstream system, Megatron-LM~\cite{megatron},
}
caches the input of each transformer layer, discards other activations, and recomputes them before backward propagation. 
\wj{
\textit{Selective Recomputation}~\cite{activation-recomputation-MLsys23} further reduces recomputation by recomputing only attention operators of each model layer.
However, both methods overlook the available GPU memory size 
and the memory requirements of the model, making them less adaptive.
\emph{Full Recomputation} easily leads to excessive recomputation time overhead.
}
Our experiments show that 
\emph{Full Recomputation} over-releases 20 GB of activations to train a 7B GPT model on 8 A100 GPUs, 
with recomputation time accounting for 10\%--30\% of total training time.
\emph{Selective Recomputation}, on the other hand, may release inadequate memory for training.

\wj{
Megatron-Uniform and Megatron-Block~\cite{megatron-ckpt} are 
two fine-grained and flexible recomputation methods in Megatron-LM.
They allow users to manually choose which layers or operators 
to recompute to avoid out-of-memory while reducing recomputation time overhead.}
However, both approaches require extensive manual efforts to find 
the suitable recomputation configuration~\cite{megatron-manual}. 
Even worse, each manual attempt requires running multiple iterations of 
training using thousands of GPUs for LLMs, incurring very high costs.

\begin{table}
	\centering
	\small
	\caption{The analysis of different activation recomputation policies.
		\label{table:comp}}
	\setlength{\tabcolsep}{1mm}{
		\resizebox{\linewidth}{!}{
			\begin{tabular}{cccc}
				\toprule
				\textbf{System}	& \textbf{Adaptivity} &\textbf{Scalability} &\textbf{Efficiency}\\
				\midrule
				\textit{Full  Recomputation}~\cite{megatron}		& \textcolor{red}{\ding{55}} & 	\textcolor{teal}{\ding{52}}	&  \textcolor{red}{\ding{55}} \\
				\midrule
				\textit{Selective Recomputation}~\cite{activation-recomputation-MLsys23}			&\textcolor{red}{\ding{55}} &\textcolor{teal}{\ding{52}} &  \textcolor{red}{\ding{55}} \\
				\midrule
				\textit{Megatron-Uniform}~\cite{megatron-ckpt}				& \textit{Manual}  &\textcolor{teal}{\ding{52}} &  \textcolor{red}{\ding{55}}\\
				\midrule
				\textit{Megatron-Block}~\cite{megatron-ckpt}		& \textit{Manual} &\textcolor{teal}{\ding{52}} & \textcolor{red}{\ding{55}} \\
				\midrule
				\textit{Checkmate}~\cite{checkmate19}		&  \textcolor{teal}{\ding{52}} &\textcolor{red}{\ding{55}} & \textcolor{red}{\ding{55}}\\
				\midrule
				\textit{AdaPipe}~\cite{AdaPipe}		& \textcolor{teal}{\ding{52}} &\textcolor{teal}{\ding{52}} & \textcolor{red}{\ding{55}}\\
				\midrule
				\pname{}		& \textcolor{teal}{\ding{52}} & \textcolor{teal}{\ding{52}} &  \textcolor{teal}{\ding{52}}\\
				\bottomrule
			\end{tabular}
		}
	}
\end{table}

\wj{
\textbf{2. Model-adaptive recomputation methods have poor scalability.}
To overcome the shortcomings of rule-based methods, Checkmate~\cite{checkmate19} 
utilizes linear programming to automatically decide which operators need recomputation 
to minimize recomputation costs. 
}
However, the search space in Checkmate increases exponentially with 
the size of the DNN models, requiring immense computational time. 
As a result, Checkmate may not provide an optimal solution within time bounds, 
limiting its scalability for training large models.
Our results show that Checkmate would take years to determine the optimal 
strategy for a 175B model, highlighting the need for faster policy-making to enable large-scale training.

\wj{
\textbf{3. All existing recomputation methods introduce significant overhead on the critical training paths.}
A recent state-of-the-art work, AdaPipe~\cite{AdaPipe}, uses dynamic programming methods 
to reduce the time required for automatically searching for optimal recomputation strategies. 
However, like all existing recomputation methods, it only starts performing recomputation 
when the released intermediate data needs to be reused. 
This introduces recomputation time on the critical training path, 
resulting in low training efficiency. 
Therefore, a method to reduce recomputation time is needed to
 improve overall training throughput.
}

\begin{figure}[t]
	\centering
	\subfigure[Communication ratio.]{
		\includegraphics[width=3.2in]{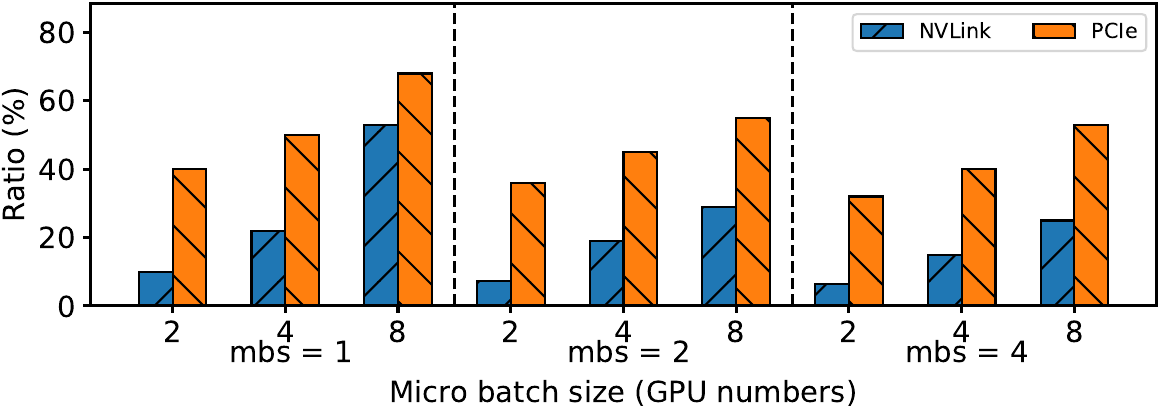}
		\label{fig:comm_ratio}
	}
	\subfigure[Memory consumption.]{
		\includegraphics[width=3.2in]{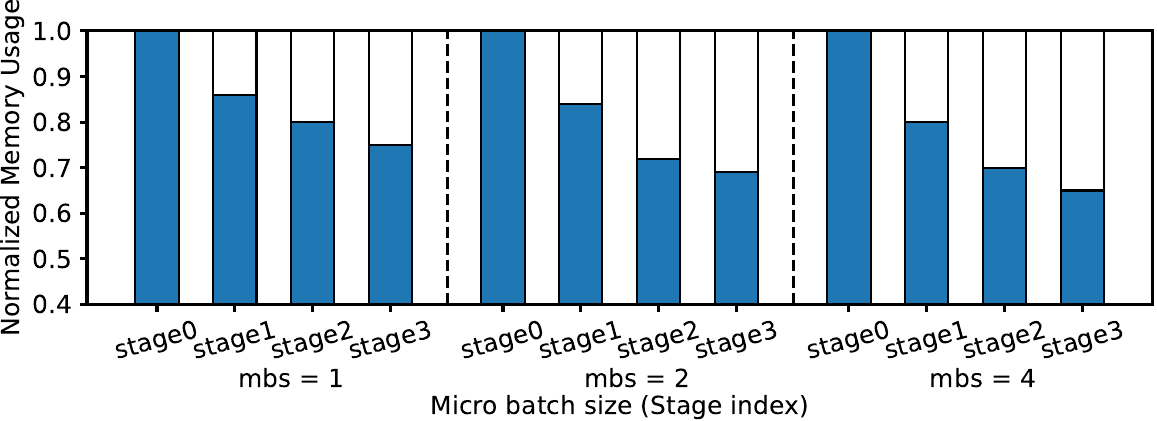}
		\label{fig:mem_ratio}	}
	\vspace{-0.1in}
	\caption{\osdi{(a) The ratio of TP communication during training. The $x$-axis represents the number of GPUs in a TP group. (b) Imbalanced stage (GPU) memory consumption (TP=2, PP=4). The memory usage is normalize with that of stage0.}
	}
	\vspace{-0.2in}
	\label{fig:opportunity}
\end{figure}

\subsection{New Opportunities}
\label{sec:opportunities}

\osdi{We experimentally find three new observations that can be used to enhance recomputation efficiency.}
Specifically, we implement a  pipeline training using both 
TP and PP to train the \osdi{7B GPT model on three micro batch size (mbs) with 1024 sequence length}. 
For PP, we divide the training process into four stages.
For TP, we use two, four, and eight GPUs for each stage. 
Experiments were conducted on NVLink- and PCIe-connected A100 GPUs.
Detailed configurations can be found in \S\ref{sec:evaluation}.

\textbf{Observation 1: existing approaches suffer from high 
communication overhead and low GPU utilization.}
\osdi{Figure~\ref{fig:comm_ratio} demonstrates that the TP communication time for
the NVLink-connected GPUs accounts for 10\%--50\% of the total training time.}
This ratio is over 70\% on PCIe-connected GPUs due to lower bandwidth.
Increasing the number of GPUs per pipeline stage reduces execution time but worsens communication bottlenecks.
Additionally, profiling reveals that SMs of GPUs 
are mostly idle during data communication, indicating low GPU utilization.

\textbf{Observation 2: GPU memory usage is imbalance across stages in training using PP.}
We observe that GPU memory is not fully utilized across GPUs and the GPU memory usage is varied across stages. For example, as shown in Figure~\ref{fig:pp}, the GPUs hosting computations in the early stages of the pipelines (e.g., GPUs in Stage0) use more memory than the others. Figure~\ref{fig:mem_ratio} shows that the highest usage of GPU memory is up to 1.5$\times$ higher than that on the GPUs with the least memory usage. This is because that activation states are generated during the forward pass for each micro batch and then kept until used by the corresponding backward pass. 
Earlier stages require storing more activation copies.
For instance, GPUs at stage 0 need to store  three copies of activation states and the GPUs at stage 3 only need to store one.

\begin{figure}
	\centering
	\includegraphics[width=3.2in]{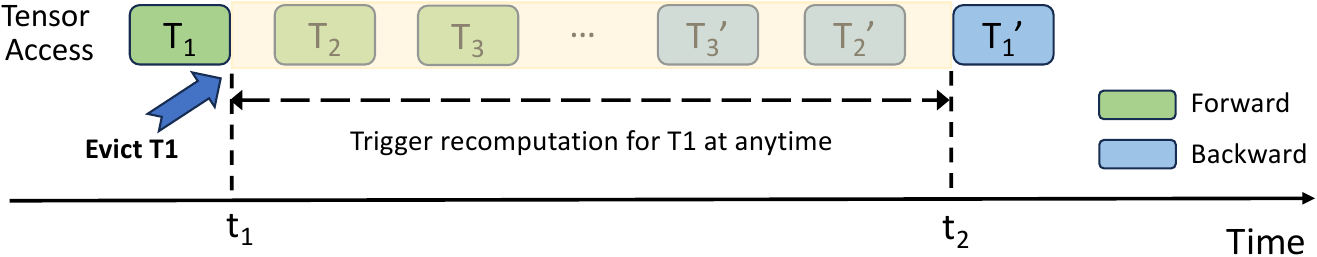}
	\caption{An example of forward, backward, and recomputation processes. $T_1$ is evicted at time $t_1$ and can be recomputed anytime between $t_1$ and $t_2$.
	}
	\label{fig:recomputation}
	\vspace{-0.2in}
\end{figure}

\textbf{Observation 3: 
Recomputation overhead is not visible until the dependent backward operation begins.
}
When the recomputation approach is used, selected activation
tensors $T$ are discarded. The backward operations $Ops$ cannot 
be executed until the selected activation tensors $T$ are recomputed.
Therefore, $Ops$ are dependent on $T$. We can schedule the
recomputation operations at any point as long as $T$ becomes
available before $Ops$ begins. Figure~\ref{fig:recomputation} illustrates this flexibility
with an example where the recomputation of $T_1$ 
can be executed anytime between $t_1$ and $t_2$.

\textbf{Opportunities.}
Current systems perform recomputation on the critical path and execute it on demand~\cite{Capuchin-ASPLOS20, megatron-ckpt}. 
Our observations highlight that we can further optimize activation 
recomputation by executing recomputation asynchronously in 
parallel with the TP communication process and selectively
discarding tensors considering their recomputation time
and the availability of idle memory space across GPUs and pipeline stages.

\section{Design of \pname{}}
\label{sec:overview}

\pname{} is designed to enable efficient memory management 
for large-model training. We have two design goals:
(1) minimizing recomputation overhead by hiding recomputation behind 
communication and (2) maximizing pipeline throughput by
model partitioning that ensures load balance across pipeline stages while accounting for recomputation time.

\begin{figure}
	\centering
	\includegraphics[width=3.2in]{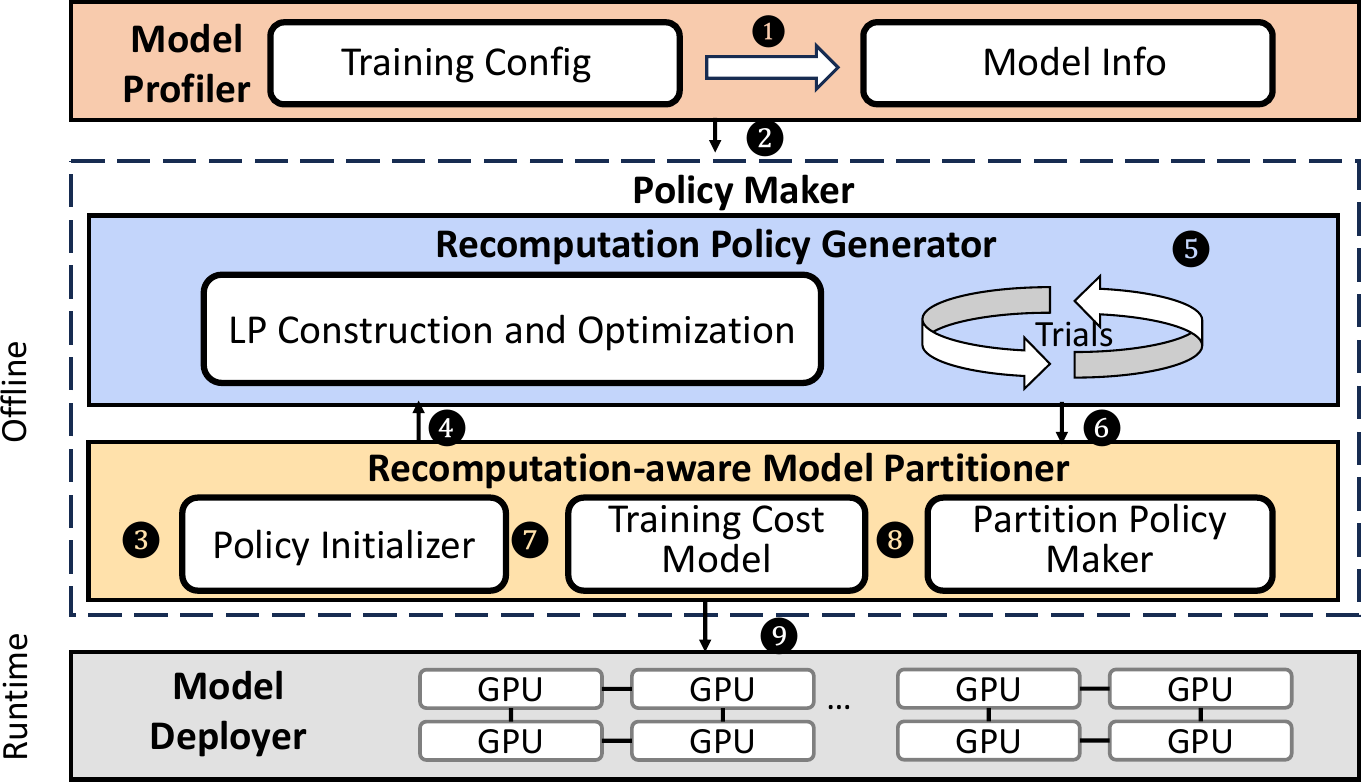}
	\caption{Overview of \pname{}.}
	\label{fig:overview}
	\vspace{-0.2in}
\end{figure}

\pname{} has three major components: 
\textit{Model Profiler}, \textit{Model Policy Maker},
and \textit{Model Deployer}. Figure~\ref{fig:overview}
shows the overview of the \pname{} software architecture. 
The functionalities of each component are described below.

\textbf{Model Profiler.} Before deploying a new model, 
we will conduct a test run using user-defined training configurations. 
These configurations include the distributed training policy (e.g., pipeline parallelism, tensor parallelism, etc.), the number of GPUs, and hyperparameters \ding{182}. 
During the test run, \pname{} collects critical model metrics including 
operator type, operator execution time, operator size, operator dependency, etc.
These metrics are
recorded in a database and serve as input for \emph{Model Policy Maker} to guide scheduling
decisions \ding{183}.
\fast{Importantly, to avoid impacting model accuracy, Lynx does not alter user-defined hyperparameters such as batch size.}


\textbf{Model Policy Maker.} It makes decisions on how to partition a model
and how to schedule a tensor recomputation considering training
throughput and load balancing among all pipeline stages.
It has two major sub-components: recomputation aware model partitioner
which generates different model partitioning schemes and recomputation policy generator
which generates a recomputation plan that minimizes recomputation
overhead for a given partitioning scheme.
\emph{Model Policy Maker} initially partitions the model and assigns them to pipeline stages \ding{184}.
This partitioning scheme is then passed to the \emph{recomputation policy generator} \ding{185} to 
determine the recomputation policy for each stage \ding{186}.
After that, the recomputation time for each stage is returned to the \emph{model 
partitioner} \ding{187}.
Then, the \emph{model partitioner} feeds the profiled forward and backward propagation times from the \emph{Model Profiler}, along with the recomputation time from the \emph{recomputation policy generator}, into the training cost model to compute the training time for each stage \ding{188}.
Finally, the \emph{Model Policy Maker} evaluates whether the pipeline achieves load balancing 
using the per-stage execution time from the \emph{model partitioner}. If not, a new partitioning scheme is generated \ding{189}, and the process repeats until load balancing is achieved.

\textbf{Model Deployer.} The \emph{Model Deployer} implements the optimal schedule determined by the \emph{Model Policy Maker}, 
utilizing deep learning frameworks to deploy the model on physical devices for training \ding{190}.

\section{Recomputation Scheduling}
\label{sec:recomp}

\begin{table*}[t]
	\centering
	\small
	\caption{\osdi{The LP constraints of \pname{}-OPT.}
		\label{tbl:opt_alg}}
	\begin{tabular}{p{3.9cm}p{4.5cm}p{8.1cm}}
		\toprule
		\textbf{Dependency Constraints.} &\textbf{Communication Constraints.} & \textbf{Memory Constraints.} \\
		\midrule
		$\textbf{(D1)} \; R_{t,i}  \leq R_{t,j} + S_{t,j}  \;  \forall t \forall i $ & $\textbf{(C1)} \;R_{t,i} = 0  \;  t, i \in COMM, t \neq i $& $\textbf{(M1)} \; U_{t,0} = M_{static} +  \sum_{i=1}^{n} M_i \times S_{t,i}   \;  \forall t$  \vspace{1pt}\\
		$\textbf{(D2)} \; S_{t,i}  \leq R_{t-1,i} + S_{t-1,i}  \;  \forall t \forall i $ &$\textbf{(C2)} \, \sum_{i=1}^{t-1} C_i \times R_{t,i} \leq C_t ,  t \in COMM$&$\textbf{(M2)}\, U_{t,i+1} = U_{t,i} + M_{i+1} \times R_{t,i+1} - \sum_{d \in DEPS(i) \cup \{i\}}  M_{d} \times F_{t,d,i} \forall t$ \\
		$\textbf{(D3)} \; R_{t,t}  = 1  \;  \forall t $&&$\textbf{(M3)} \;	F_{t,d,i} = R_{t,i} \times (1 - S_{t+1,d})\times \prod_{j\in USER(d),j>i} (1-R_{t,j})$ \vspace{1pt}\\
		$\textbf{(D4)} \; S_{1,i}  = 0  \;  \forall i$&&$\textbf{(M4)} \;U_{t,i} \leq M_{budget} \quad \forall t \forall i$ \vspace{1pt}\\
		\bottomrule
	\end{tabular}
	\vspace{-0.2in}
\end{table*}

\osdi{
Our goal is to develop a recomputation policy that maximizes training throughput while preventing out-of-memory issues.
This requires addressing key challenges:
(1) deciding which tensors to recompute, 
(2) determining whether recomputation is on the critical path or overlaps with communication,
(3) identifying the communication phase for the recomputation to overlap with,
and (4) ensuring the policy is yielded within an acceptable time.
Given the NP-hard nature of recomputation scheduling, we use LP formulations to find solutions.
To determine the upper bound of achievable throughput, we design an optimal LP called \opt{} in \S\ref{sec:optimal}. 
To address the vast search space of \opt{},
we introduce \heu{}, a heuristic approach that provides near-optimal solutions within a reasonable time  in \S\ref{sec:heuristic}.
While this work focuses on homogeneous cluster, our formulation can be extended to heterogeneous GPU clusters,
which we plan to explore in future work.
}

\begin{table}[t]
	\centering
	\small
	\caption{\osdi{Variables used in \pname{}.}
		\label{tbl:variables_opt}}
			\setlength{\tabcolsep}{0.5mm}{
			\resizebox{\linewidth}{!}{
	\begin{tabular}{ll}
		\toprule
		\textbf{Variables} &\textbf{Description} \\
		$OP_{n}$ & N training operators\\
		$Phase_{n}$ & N training execution phase. $N$ operators correspond to $N$  
		phases\\
		$COMM$ & Sets of communication operators (e.g., all-reduce).\\
		$C_{i}$ & Computation time of $OP_{i}$\\
		$M_{i}$ & Output memory of $OP_{i}$\\
		$M_{budget}$&The peak memory is limited by the GPU memory \\
		$M_{static}$& Static memory including  parameters, gradients, and optimizer states\\
		$R_{t,i}$ & Whether $OP_i$ is computed at $Phase_{t}$\\
		$S_{t,i}$& The output of $OP_i$ is retained in GPUs from $Phase_{t-1}$ to $Phase_{t}$\\
		$U_{t,i}$&$U_{t,i}$ $\in \mathbb{R}^+$. The memory used after computing $OP_i$ in $Phase_t$\\
		$F_{t,d,i}$ & Whether the output of $OP_d$ can be freed in $Phase_t$ after $OP_i$ is computed
		\\
		\bottomrule
	\end{tabular}
}}
	\vspace{-0.2in}
\end{table}

\subsection{Optimal Recomputation Scheduling}
\label{sec:optimal}

\osdi{

In this section, we present the \pname{}-OPT algorithm and summarize the challenges of operationalizing \pname{}-OPT.
Table~\ref{tbl:opt_alg} and Table~\ref{tbl:variables_opt} summarizes all constraints and used variables.

}

\textbf{Problem definition.} 
\fast{The DNN model comprises $N$ operators ($OP_{n}$) that perform training operations based on the model topology.}
$OP_i$ must be executed at $Phase_i$.
Other operators can also be performed at $Phase_i$ for tensor recomputation.
Whether $OP_i$ can be executed depends on whether 
the result of its preceding dependencies $OP_j$ (where $j < i$) 
have been available in the device.

\textbf{Objective.} The output of each operator can be either saved 
in GPUs or recomputed. Our objective is to minimize the end-to-end training time 
along the critical path including forward time, 
backward time, and recomputation overhead.
In other words, we need to minimize the total computation time for 
all operators minus the overlapped recomputation time during 
communication:

\begin{equation}
\label{eq:objective}
{\small
\begin{aligned} 
& \underset{R}{\text{minimize}} & & \sum_{t=1}^{n} \sum_{i=1}^{t} C_i \times R_{t,i} - \sum_{t \in COMM} \sum_{i=1}^{t-1} C_i \times R_{t,i} \\
& \text{subject to} && \text{Constraints in Table~\ref{tbl:opt_alg}}\\
\end{aligned}
}
\end{equation}

\textbf{Dependency constraints.} Constraint D1 and D2 ensure that $OP_i$ is 
computed in $Phase_t$ only if all dependencies (i.e., outputs of $OP_j$) of $OP_i$ are available. 
In D1, 
the execution of $OP_i$ requires that
$OP_j$  is either executed at $Phase_t$ ($R_{t,j}$) or its output was generated before $Phase_t$ ($S_{t,j}$).
According to our definitions, $OP_i$  must execute at $Phase_i$, as shown in D3. 
In the first phase of training,
D4 specifies that no tensor are initially in memory. 


\textbf{Communication constraints.} 
 \pname{} is the first work to consider how to overlap recomputation with communication.
Overlapping recomputation is challenging because
recomputation also has communication operators.
These communication operations cannot overlap with the communication involved 
in forward or backward training due to bandwidth conflicts~\cite{MegaScale-arxiv24}. 
We define C1 to formulate this constraint.
Additionally, we must prevent the overlapped recomputation time from exceeding the 
communication time, otherwise it may induce memory pressure for preloading 
the intermediate data on the device without substantial performance gains (C2).


\textbf{Memory constraints.} 
For each phase, in addition to  the fixed memory consumption ($M_{static}$), three factors dynamically impact memory usage: (1) checkpointed tensors stored in the device (determined by $S$); (2) tensors generated during training (determined by $R$); and (3) memory reduction resulting from freed tensors.


We initialize the memory usage in M1
and recursively evaluate it (in M2) for all operations in  $Phase_t$, considering newly generated tensors and freed memory.
In M2, $DEPS(i)$ represent the dependent operators (parents) of $OP_i$.
We define $F_{t,d,i}$ in M3, where  $USER(d)$ represents operators dependent on  $OP_d$ (children of  $OP_d$):


The output of $OP_d$ can be discarded after the execution of $OP_i$ if three conditions are met: (1) $OP_i$ is executed in $Phase_t$, (2) $OP_d$ is not checkpointed for $Phase_{t+1}$, and (3) $OP_d$'s children are not executed in $Phase_t$.
We apply De Morgan's law and intersection interchange techniques from Checkmate~\cite{checkmate19} to linearize this equation, omitting details for brevity.
Finally, memory usage for any phase must remain within the device constraint, as described in M4.



\osdi{
\textbf{Challenges of operationalizing \pname{}-OPT.}
While optimal recomputation identified by \opt{} provides an upper bound on the  training performance, it is impractical for models with  a large number of  layers to use \opt{},
due to the extensive search space resulting from forward and backward operators.
\wj{
Despite using optimization techniques in~\cite{STR,checkmate19},
\textit{\pname{}-OPT} takes 14 hours to generate policies even for relatively small models like GPT-300M (\S\ref{eval:heu}).}
As model size grows, search time increases exponentially—potentially requiring months or years for models with tens of billions of parameters.
Therefore, a more practical algorithm is needed to generate policies within an acceptable time bound.
}



\begin{figure*}
	\centering
	\includegraphics[width=7in]{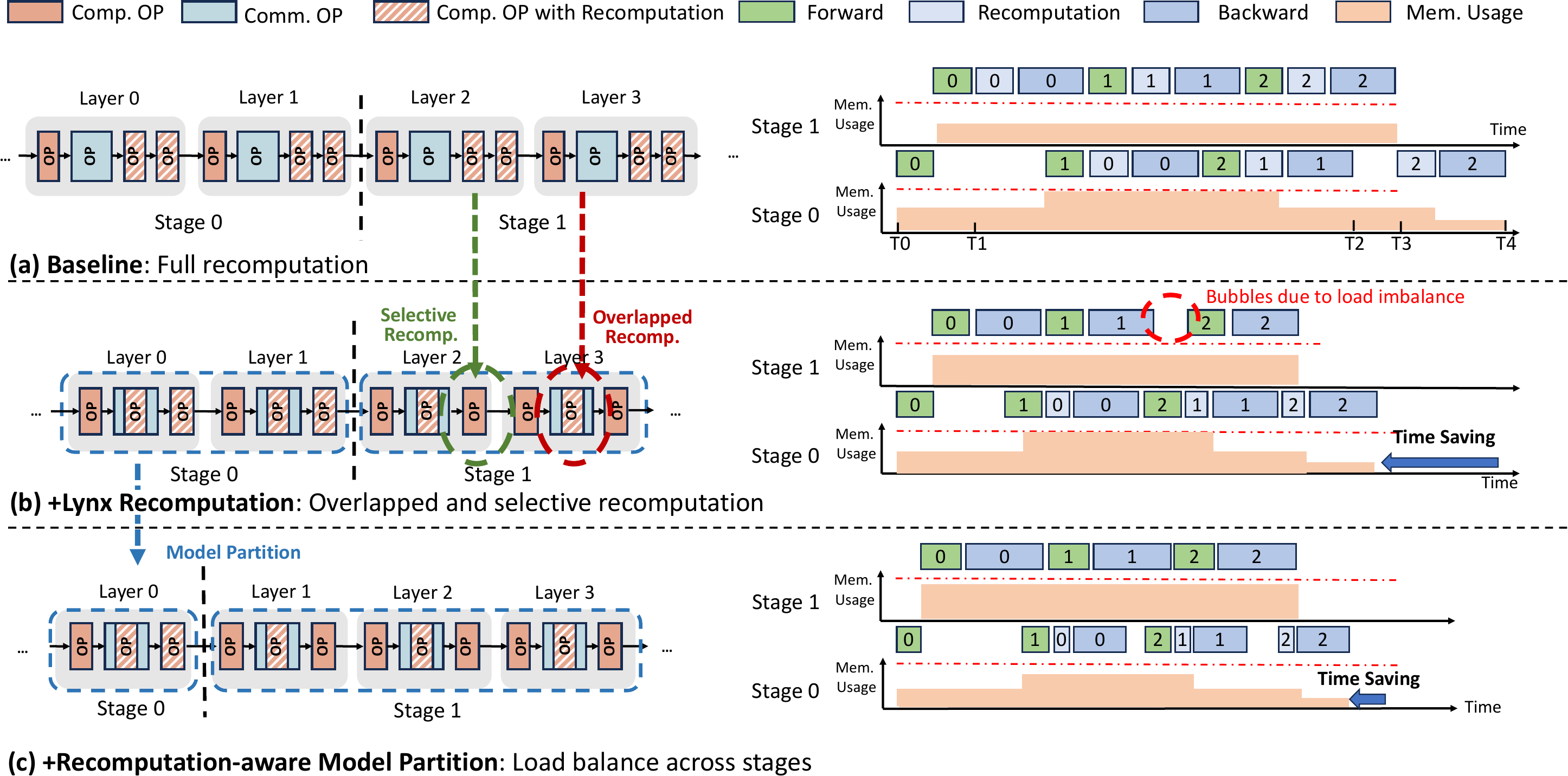}
	\vspace{-0.1in}
	\caption{\wj{
			The left part illustrates the recomputation and model partitioning policy in \pname{}, while the right part shows the time and memory usage of the two stages with 3 microbatches (0 to 2). 
			Comp. OP and Comm. OP represent the computation and communication operators, respectively. 
			The output of shaded Comp. OPs is released from GPU after the forward pass and regenerated through recomputation during backward. 
			The red dashed line on the right figure indicates the GPU memory budget.
	}}
	\vspace{-0.2in}
	\label{fig:details}
	
\end{figure*}
\subsection{Heuristic Recomputation Scheduling}
\label{sec:heuristic}

In this section, we describe a heuristic-based recomputation scheduling
approach, \pname{}-HEU, to reduce the search time while achieving close-to-optimal training performance.


\textbf{Key observation of identical structures.} Large DNN models consist of 
multiple identical structures. For example, as shown in Figure~\ref{fig:details}(a), the pipeline parallelism has three fixed training procedures~\cite{MegaScale-arxiv24}, including 
\textit{warm-up} (T0--T1), \textit{steady} (T1--T2), and \textit{cool-down} (T2--T4). Each
procedure contains repeated training structures. Specifically, 
(1) there are several identical forward passes during \emph{warm-up}. 
(2) During \emph{steady}, each worker executes 
the pattern of one forward propagation followed by one backward propagation (i.e., 1F1B).
(3) During \emph{cool-down}, workers perform the repeated pattern of 
one synchronization stall followed by one recomputation and backward pass. 
Similarly, large-scale models, such as GPT~\cite{attention}, consist of
numerous identical layers, 
like transformer layers (e.g., Layers 0--3 in Figure~\ref{fig:details}), which exhibit similar 
GPU memory footprints and computing times.

\wj{
\textbf{Key idea.}
We find that the local optimal recomputation policy for a single structure/layer can be applied to other identical structures/layers without triggering 
the search in the global space. For example, as shown in Figure~\ref{fig:details}(a), there are many repeated 1F1B training patterns 
in the \emph{steady} stage (T1--T2), with each 1F1B training period involving multiple identical transform layers. 
Therefore, we can establish a policy for a single transform layer and apply this policy across layers and patterns.}
We formulate the problem as a linear program (LP), accounting for operator dependencies, overlapped recomputation and communication constraints, and device memory limitations.

\textbf{Problem definition. } 
A single basic layer (e.g., a transformer layer) consists of $N$ operators 
($OP_1$, ..., $OP_{n}$). For each layer, there are four communication phases 
that can be used for hiding recomputation time, including two forward 
communication phases (named $Phase_{1}$ and $Phase_{2}$) and 
two backward communication phases (named $Phase_{3}$ and $Phase_{4}$) as shown in Figure~\ref{fig:tp}.
In addition, if overlapping is not feasible, we can always 
execute the recomputation on-demand in the critical path ($Phase_{5}$).
The definitions of $R_{t,i}$, $M_{i}$, and $C_{i}$, $COMM$ are
the same as in Table~\ref{tbl:variables_opt}. Boolean $S_i$ denotes whether the 
output of $OP_i$ will be retained in GPUs permanently.
Besides, the forward passes of \emph{warm-up} and \emph{steady} share identical
tensor retention and recomputation policies in our design.

\textbf{Objective.} Our objective is to minimize the recomputation time in the critical path for a basic model layer. 
In Equation~\ref{eq:objective2},
$(1-S_i)=1$ indicates $OP_i$ is recomputed, and $R_{5,i}=1$ represents $OP_i$ is recomputed in the critical path.

\begin{equation}
\label{eq:objective2}
{\small
\begin{aligned} 
& \underset{S,R}{\text{minimize}} & & \sum_{i=1}^{n} (1-S_i)  \times R_{5,i} \times C_i \\
\end{aligned}
}
\end{equation}
\vspace{-0.1in}

\textbf{Dependency constraints.} We constraint each recomputation operator to be executed only once in Equation~\ref{eq:cons11}. 
Whether $OP_i$ can be executed in $Phase_t$ depends on whether $OP_j$ is computed before $Phase_t$ or has been stored in the GPU, where $OP_j$ is the preceding dependent operator of $OP_i$, as illustrated in Equation~\ref{eq:cons12}. 
{\small
	\begin{align}
		\sum_{t=1}^{5} R_{t,i} = 1 \quad  \forall i \label{eq:cons11} 
	\end{align}
}
\vspace{-0.2in}
{\small
	\begin{align}
		R_{t,i} \leq \sum_{t'=1}^{t} R_{t',j} +S_j \quad  t \in [1,5],  \forall i \label{eq:cons12}
	\end{align}
}



\textbf{Communication constraints.} 
We need to ensure that the overlapped recomputation time does not exceed the communication time (Equation~\ref{eq:cons13}), and communication operators should not be invoked during the communication process (Equation~\ref{eq:cons14}).
\fast{This is because (1) limiting overlapped recomputation to stay within communication time reduces search time while maintaining performance, and (2) preventing concurrent communication avoids network contention,  leading to more accurate performance predictions for policy maker and effective scheduling algorithms.}

\begin{figure}
	\centering
	\includegraphics[width=3.3in]{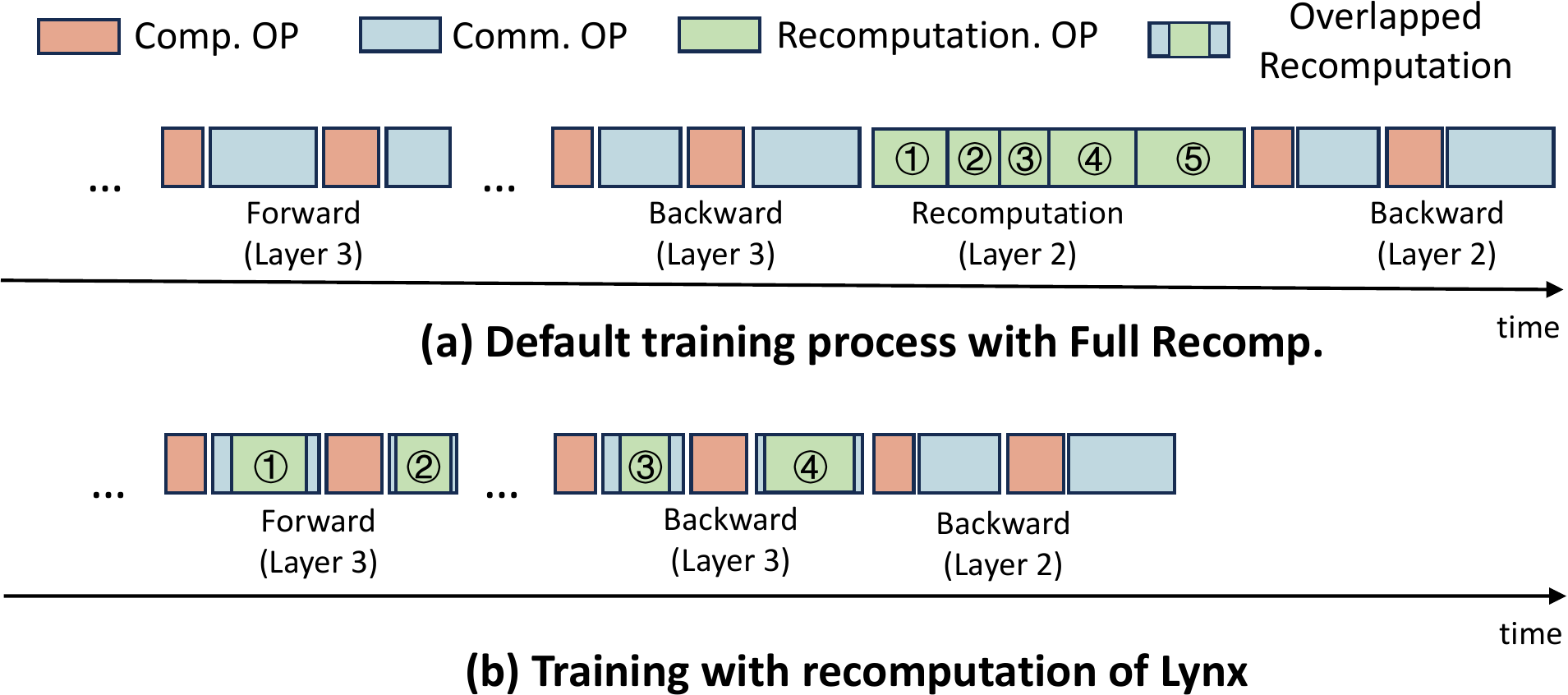}
	\caption{\osdi{The training details of 1F1B  correspond to  Stage 1 in Figure~\ref{fig:details}(a) and (b).
}}
	\label{fig:overlap_re}
	\vspace{-0.1in}
\end{figure}

\begin{equation} 
	\label{eq:cons13}
	\sum_{i=1}^{n} (1-S_i) \times R_{t,i} \times C_i \leq CTime_t \quad t \in [1,4]
\end{equation}

where $CTime_1$ and $CTime_2$ represent two forward communication time, and $CTime_3$ and $CTime_4$ represent two backward communication time, respectively.

\begin{equation} 
	\label{eq:cons14}
	R_{t,i} = 0 \quad t \in [1,4] \; i\in COMM 
\end{equation}

\textbf{Memory constraints.} 
We need to ensure the peak memory usage is smaller than the GPU memory size ($M_{budget}$).
Since unnecessary tensors are gradually released during backward propagation, the peak memory 
usage occurs before the first backward propagation begins~\cite{AntMan-OSDI20}.
Therefore, we define the peak memory usage as Equation~\ref{eq:cons15}.
Specifically, the peak memory comprises the fixed memory ($M_{static}$), tensors 
($M_{fwd}$) residing in the GPU after forward propagations before the first 
backward propagation, and tensors generated during the forward communication ($M_{fwd\_comm}$). 

\begin{equation} 
	\label{eq:cons15}
	M_{static} + M_{fwd} + M_{fwd\_comm}  \leq M_{budget}
\end{equation}

$M_{fwd}$ is formulated in Equation~\ref{eq:cons16}, where $N_{layer}$ denotes the number of transformer layers in the DNN model, and $N_{batch}$ represents the number of forward pass before  the first backward propagation (e.g., Stage0 has 4 forward passes in Figure~\ref{fig:pp}).
We define $S_{n} = 1$ to store the output of $OP_n$  in GPU as the checkpoint.


\begin{align}
M_{fwd} = (N_{layer} \times \sum_{i=1}^{n} S_i \times M_i) \times N_{batch} \label{eq:cons16}
\end{align}

In our design, recomputation is not overlapped with communication during the \emph{warm-up} phase, as no recomputation operations occur in this phase.
Therefore, we only calculate the size of data generated during forward communication for a single forward batch in the \emph{steady} phase:

{\small
\begin{equation} 
	\label{eq:cons17}
	M_{fwd\_comm} = N_{layer} \times \sum_{i=1}^{n} (1-S_i) \times (R_{1,i} + R_{2,i}) \times M_i
\end{equation}
}


\textbf{Optimizations.}
First, in the last pipeline stage (e.g., Stage3 in Figure~\ref{fig:pp}), 
it is unnecessary to overlap recomputation in the forward communication 
because recomputation will be immediately executed after discarding the 
corresponding tensors.  
In this scenario, we only consider 3 phases defined in LP: two backward communications 
and the critical path for on-demand recomputation.
When modeling the LP for the last pipeline stage, we exclude  $M_{fwd\_comm}$ in the memory constraint.

\osdi{Second, the recomputation scheduling during \emph{cool-down} can be
further improved. The training in \emph{cool-down} incurs
many synchronization stalls (T2--T3) in Figure~\ref{fig:details}(a). \pname{} further uses the synchronization
stalls for hiding recomputation overhead when all the dependent
tensors are on the same GPU and sufficient GPU memory is available.
For example, in Figure~\ref{fig:details}(b), \pname{} parallelizes the recomputation of Batch 2 in Stage 0 with preceding synchronization stalls, further improving training efficiency.
}

\osdi{\textbf{Example.}
Figure~\ref{fig:details}(a) and (b) show the recomputation cases with Full recomputation and \pname{}-HEU. 
Full recomputation disregards unused GPU memory, leading to excessive recomputation.
In contrast, since GPUs in Stage 1 has ample memory space, \pname{}-HEU stores some activations on the GPU to reduce recomputation overhead (Selective Recomp. in Figure~\ref{fig:details}(b)). 
Additionally, \pname{}-HEU overlaps part of the recomputation with communication, further reducing training time (Overlapped Recomp.).
\wj{
Figure~\ref{fig:overlap_re} illustrates the details. Assume that the first shaded OP of Layer 2 in Figure~\ref{fig:details}(a) corresponds to four finer-grained operators \ding{172}--\ding{175}, and the second shaded operator corresponds to a single operator \ding{176}. In \pname{}-HEU,
}
\ding{172}--\ding{173}  are overlapped with communication during the forward pass of Layer 3 in the previous batch; \ding{174}--\ding{175}  are overlapped with communication during backward pass of Layer 3 in the current batch; \ding{176}  is avoided entirely by storing its activation on the GPU.
Thus, we eliminate all recomputation overhead in Stage 1 of Figure~\ref{fig:details}(b).
}



\textbf{Search time.}
\pname{}-HEU significantly 
reduces the search space, requiring less than seconds to find an optimal 
policy in our evaluation (even for the very large model of 175B). 
More details are shown in \S\ref{eval:overhead}.

\section{Recomputation-Aware Model Partitioning}
\label{sec:partition}

\begin{figure}
	\centering
	\includegraphics[width=3.3in]{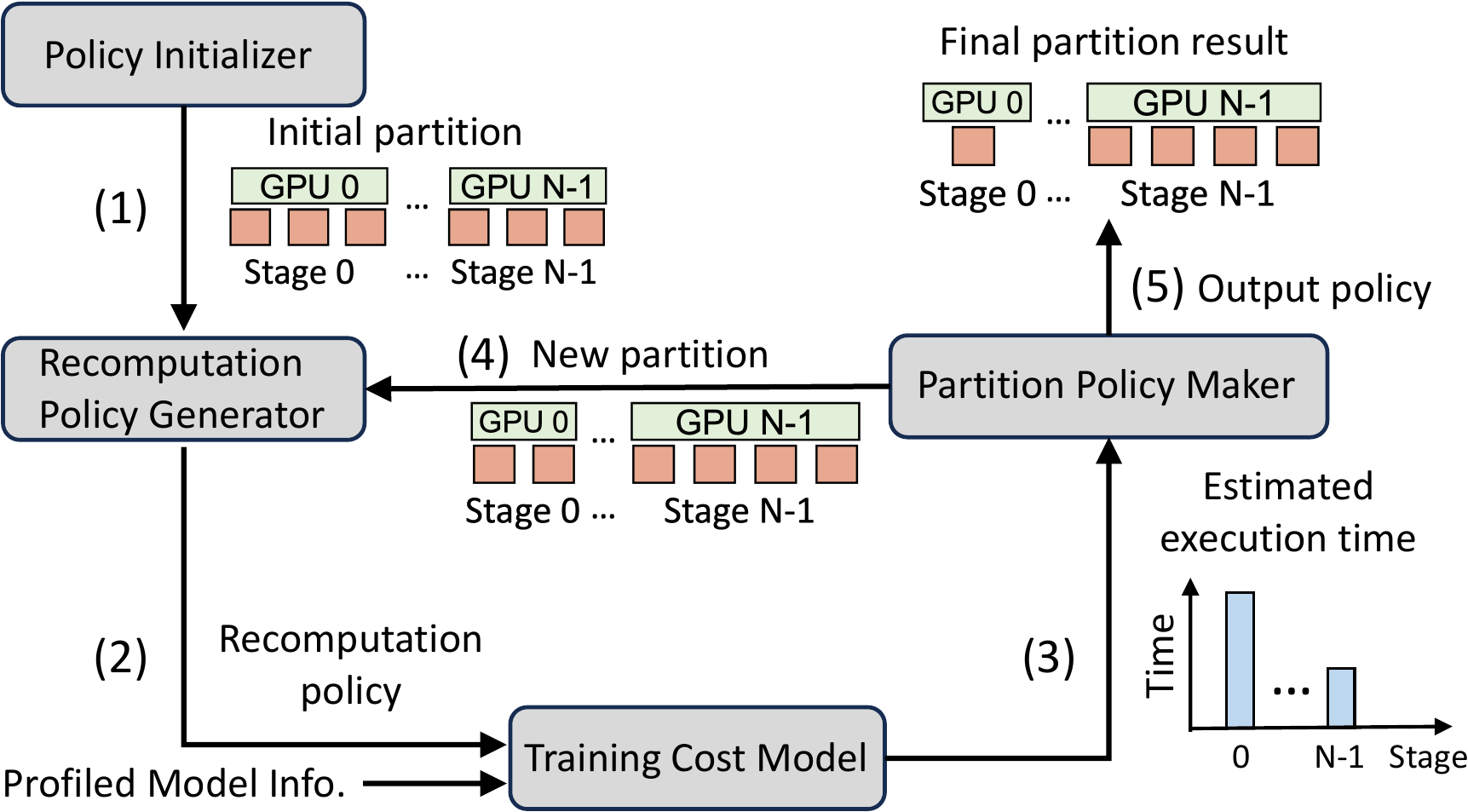}
	\caption{\wj{The recomputation-aware model partitioning approach. Each red rectangle represents a layer of the model.}}
	\label{fig:recomp-aware-part}
	\vspace{-0.2in}
\end{figure}

In this section, we describe a model partitioning approach
that can achieve computation times balancing among pipeline stages when recomputation
is overlapped with communication. 
\fast{It is not independent of the recomputation policy, 
	as model partitioning is related to the recomputation scheduling algorithm.}
We use a greedy algorithm in the search of a partitioning policy as shown in \wj{
Figure~\ref{fig:recomp-aware-part}}.

\wj{
\textbf{Key idea.}
We aim to iteratively reduce the number of layers in the slowest stage and increase the number of layers in the fastest stage until the execution times across stages are as equal as possible.
Specifically, it consists of four steps: (1) The Policy Initializer generates an initial model partitioning scheme, where each stage has approximately the same number of layers; (2) Based on the current partitioning scheme, the Recomputation Policy Generator uses heuristic recomputation scheduling~\ref{sec:heuristic} to find the near-optimal recomputation policy; (3) The Training Cost Model estimates the total time for each stage based on the recomputation policy, leveraging information collected by the Model Profiler; (4) The Partition Policy Maker generates a new model partitioning scheme by reducing one layer from the slowest stage and adding one layer to the fastest stage. 
Then, steps (2) and (3) are re-executed to evaluate each stage's  execution time.}
If the new partitioning scheme is valid (i.e., no out-of-memory errors) 
and the longest stage of the new partitioning scheme
is shorter than the current longest stage, then the new partitioning scheme is adopted.
\wj{
Repeat steps (2), (3), and (4) until the partitioning scheme does not change compared to the last iteration, yielding the output policy (5).
}

\wj{\textbf{Example.}
Figure~\ref{fig:details}(b) shows the case without recomputation-aware model partitioning, where each stage has two layers.
Stage 0 has the highest storage pressure, requiring recomputation of both shaded OPs, with only one overlapped with communication.
In contrast, in Stage 1, only one shaded OP needs recomputation and can be fully overlapped. 
This causes Stage 0 to take longer than Stage 1 due to the additional recomputation time on the critical training path, leading to a pipeline bubble.
In contrast, Figure~\ref{fig:details}(c) shows that after enabling recomputation-aware model partitioning, Stage 0 is assigned one layer and Stage 1 is assigned three layers. After readjusting the recomputation policy, the total times for both stages are approximately balanced, improving pipeline efficiency and reducing total training time.
}

\osdi{
\section{Implementation}
\label{sec:implementation}


\textbf{Model Profiler.}
It collects model metrics before training and addresses two challenges.
(1) Profiling the full model risks out-of-memory (OOM) issues and high computational costs.
To address this,  \pname{} profiles only a single representative layer  instead of a group of similar layers, leveraging the repetitive structures common in large-scale models.
(2) Modeling hundreds of operators individually for the linear program (LP) formulation is inefficient. To reduce search overhead, smaller operators (e.g., add, get shape, and transpose) are grouped into a single unit, while major operators (e.g., Matmul and Fused Layernorm) remain as individual scheduling units.

\textbf{Policy Maker.}
It supports any hybrid parallelism policy and applies search algorithms based on profiled metrics, allowing \pname{} to find the optimal partitioning and recomputation policy for each PP stage.
To improve practicality,
we implement the policy search algorithm using the Gurobi optimizer~\cite{gurobi} and integrate it into the profiling interface, streamlining the process by combining profiling and policy making.

\textbf{Model Deployer.}
It supports two training frameworks, Megatron-LM~\cite{megatron} for NVIDIA GPUs and MindSpeed~\cite{MindSpeed} for Ascend NPUs.
Both frameworks provide basic interfaces for computation graph partitioning.
We implement \pname{}'s partitioning using these interfaces, define custom overlapped and selective recomputation, and modify the decoder layer to support our recomputation policies.

}

\section{Evaluation}
\label{sec:evaluation}

\subsection{Experimental Setup}
\label{sec:setup}



\begin{figure*}[t]
	\centering
	\subfigure[NVLink.]{
		\includegraphics[width=3.3in, height=1.2in]{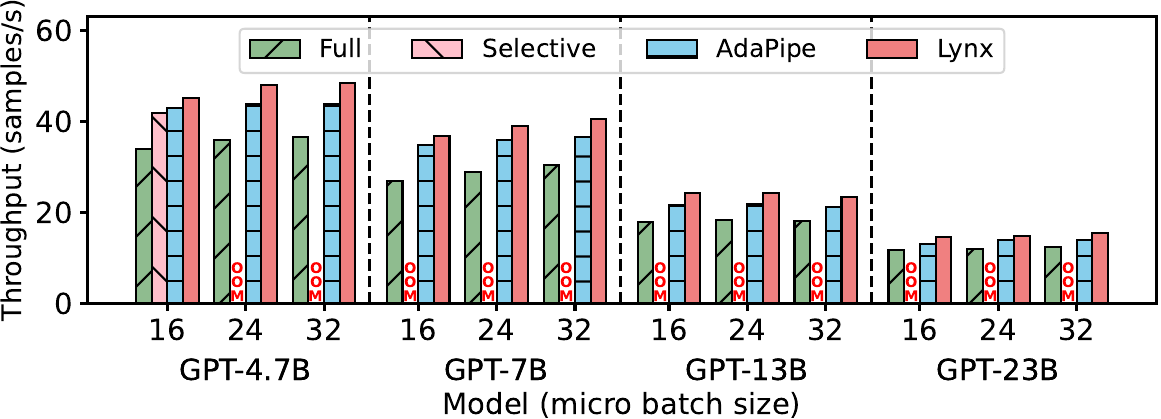}
		\label{fig:general_nvlink}
	}
	\subfigure[PCIe.]{
		\includegraphics[width=3.3in, height=1.2in]{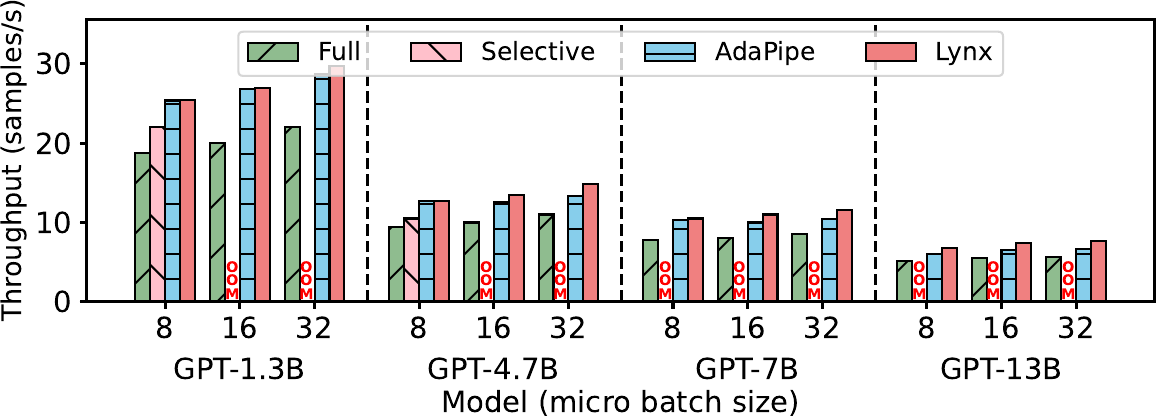}
		\label{fig:general_pcie}	}
	\vspace{-0.1in}
	\caption{\osdi{Overall training throughput of different recomputation policies across five models and two GPU clusters.  We omit displaying evaluation results that encounter out-of-memory issues.}
	}
	\vspace{-0.1in}
	\label{fig:overall}
\end{figure*}

\begin{figure*}[]
	\centering
	\subfigure[Recomputation overhead comparison.]{
		\includegraphics[width=3.3in,height=1.1in]{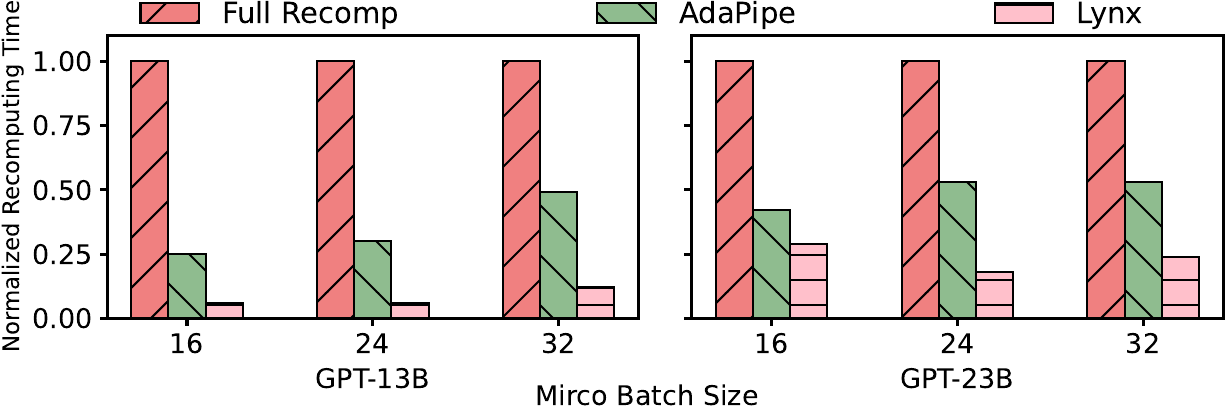}
		\label{fig:recomputation_overhead}
	}
	\subfigure[Recomputation details.]{
		\includegraphics[width=3.3in,height=1.1in]{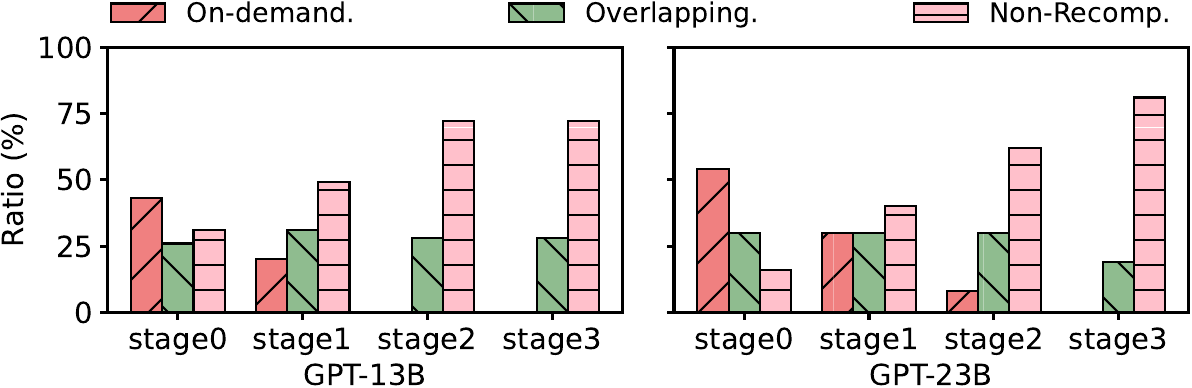}
		\label{fig:recomputation_breakdown}	}
	\vspace{-0.1in}
	\caption{\osdi{(a)  The time overhead is normalized to that of Full Recomputation. (b) Time breakdown of \pname{} recomputation of four pipeline stages with 32 micro batch size (mbs).}
	}
	\vspace{-0.2in}
	\label{fig:recompute}
\end{figure*}

\textbf{Clusters.}
\wj{We conduct experiments primarily on two NVIDIA clusters with different GPUs and network bandwidths.}
The \emph{NVIDIA-NVLink cluster} consists of four nodes, each with 256GB DRAM, two Intel Xeon Gold 6130 CPUs and four NVIDIA A100-SXM 40GB GPUs interconnected via NVLink (600 GB/s bidirectional bandwidth).
The \emph{NVIDIA-PCIe cluster} also consists of four nodes, each equipped with 128GB DRAM, two Intel Xeon Gold 5318Y CPUs, and two NVIDIA A100-PCIe 40GB GPUs with PCIe 4.0 (64 GB/s bidirectional bandwidth). All nodes are connected via ConnectX-5 Infiniband.

Additionally, we evaluate \pname{} on an \textit{Ascend NPU cluster} (\cref{eval:sensitivity}). It consists of four nodes,  each with eight Ascend 910 32GB Accelerators (NPUs), 192 CPU cores in four sockets, and 512GB memory. The eight NPUs are installed on two NPU boards in each node, and the four NPUs on each board are fully meshed via 30 GB/s links in all directions.
All nodes are connected via a 100 Gbps NIC for inter-node communication.

\wj{
\textbf{Baselines.}
We compare \pname{} with 
the following systems:
(1) \textit{Full Recomputation}~\cite{megatron}: It releases all intermediate data in each layer and recomputes all model layer before backward. For model partitioning, it balances the number of model parameters on each pipeline stage~\cite{Deepspeed-Partition}. We name this default partitioning approach as the \textit{dp-partitioning}.
(2) \textit{Selective Recomputation}~\cite{activation-recomputation-MLsys23}: It only recomputes the attention operators within each layer and also adopts the \emph{dp-partition}.
(3) \textit{AdaPipe}~\cite{AdaPipe}: AdaPipe is the state-of-the-art model-adaptive recomputation system. It automatically determines the recomputation and model partitioning strategy through a dynamic programming algorithm.
All these systems expose recomputation time along the critical computation path.
In contrast, \pname{} parallelizes recomputation time with communication time using the heuristic recomputation scheduling and applies a recomputation-aware model partitioning strategy.

Besides, we compare \pname{} with \pname{}-OPT to demonstrate \pname{}'s superiority in balancing recomputation policy search overhead and model training performance in \S\ref{eval:heu}.
We do not compare \pname{} with Megatron-Block and Megatron-Uniform, as \textit{AdaPipe} already outperforms them.

\textbf{Workloads.}
We use six GPT~\cite{gpt2} models of varying scales: GPT-300M, GPT-1.3B, GPT-4.7B, GPT-7B, GPT-13B, and GPT-23B.
They have varying attention heads, hidden dimensions, and numbers of layers, as specified in the official documentation~\cite{gpt3}. 
If not specified, the sequence length is set to 1024.
All models are trained on the representative WikiText2 dataset~\cite{WikiText2} using mixed-precision training, following the approach outlined in related work~\cite{megatron-wiki}.
}

\subsection{Overall Performance}
\label{eval:overall}

\wj{
Figure~\ref{fig:overall} shows the model training throughput results 
for different models with varying micro batch sizes (ranging from 8 to 32) 
across two clusters. 
The micro batch size refers to the number of training samples per GPU. 
We have the following five observations.

First, \pname{} outperforms others, with up to 1.37$\times$, 1.2$\times$, and 1.18$\times$ throughput gains over Full Recomputation, Selective Recomputation, and AdaPipe, respectively, highlighting its effectiveness.
Second, \pname{} achieves greater average speedup on the NVIDIA-PCIe cluster (1.35$\times$) than on the NVLink cluster (1.3$\times$) compared to Full Recomputation, as slower PCIe bandwidth allows more recomputation to overlap with communication.
Third, \pname{}'s speedup varies by model scale and micro batch size, as these factors influence \pname{}'s recomputation and partitioning strategies.
Fourth, \pname{} outperforms AdaPipe by up to 1.18$\times$ and 1.2$\times$ on NVLink and PCIe clusters, respectively. This is because \pname{} can overlap recomputation with communication, further reducing recomputation overhead (details in \S\ref{eval:recomp}).
Fifth, Selective Recomputation faces out-of-memory issues with large models or micro batch sizes due to insufficient memory release, unlike \pname{} which adapts recomputation policies to GPU memory.

}

\subsection{Breakdown Analysis}

\subsubsection{Effectiveness of Recomputation Policy}
\label{eval:recomp}

\textbf{Recomputation time comparison.}
We use the \textit{dp-partitioning} in all the 
experiments, ensuring an even distribution of model parameters 
across each pipeline stage.
\osdi{
Due to space constraints, we present results for GPT-13B and GPT-23B models on the NVIDIA-NVLink cluster only.
Similar trends are observed for other models and configurations. We exclude all selective recomputation results due to OOM issues.
 Figure~\ref{fig:recomputation_overhead} shows the normalized recomputation time on the critical path. We observe that \pname{} reduces recomputation time by 71\%--94\% and 31\%--80\% compared to Full Recomputation, and AdaPipe, respectively. This is because \pname{} selects appropriate layers for recomputation and hides recomputation time within communication.

\textbf{Recomputation operator ratio.}
Figure~\ref{fig:recomputation_breakdown} shows the ratio of 
recomputation operators on the critical path (denoted as \textit{On-demand}), 
recomputation operators run in parallel with communication (denoted as \textit{Overlapping}), 
and non-recomputation operators (denoted as \textit{Non-Recomp.}) in \pname{}. 
\pname{} achieves up to 31\% and 30\% recomputation-communication overlap on the 13B and 23B models, respectively, with a uniform proportion across stages.
We also observe that \pname{} effectively reduces more recomputation overhead in the later pipeline stages.
For example, it eliminates all recomputation overhead in stage 2 and stage 3 for the 13B model, while reducing it by only 57\% and 80\% in stage 0 and stage 1, respectively.
This is because training in the earlier stages consumes more GPU memory, making it hard to fully hide recomputation within communication.}

\subsubsection{Effectivness of Model Partitioning}
\label{eval:partition}
\vspace{-0.1in}

\begin{figure}
	\centering
	\includegraphics[width=3.2in,height=1.1in]{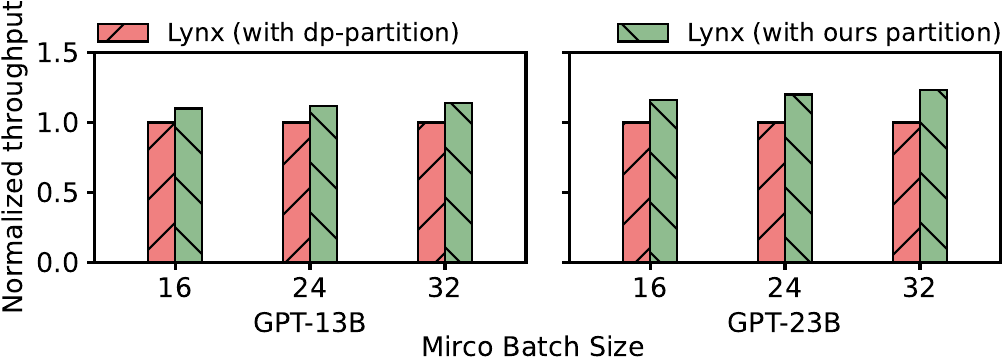}
	\caption{Training throughputs with different model partitioning schemes.
		The throughput (samples/s) is normalized to that of Lynx with dp-partition.}
	\label{fig:partition_eval}
\end{figure}

\begin{figure}
	\centering
	\includegraphics[width=3.3in, height=1.1in]{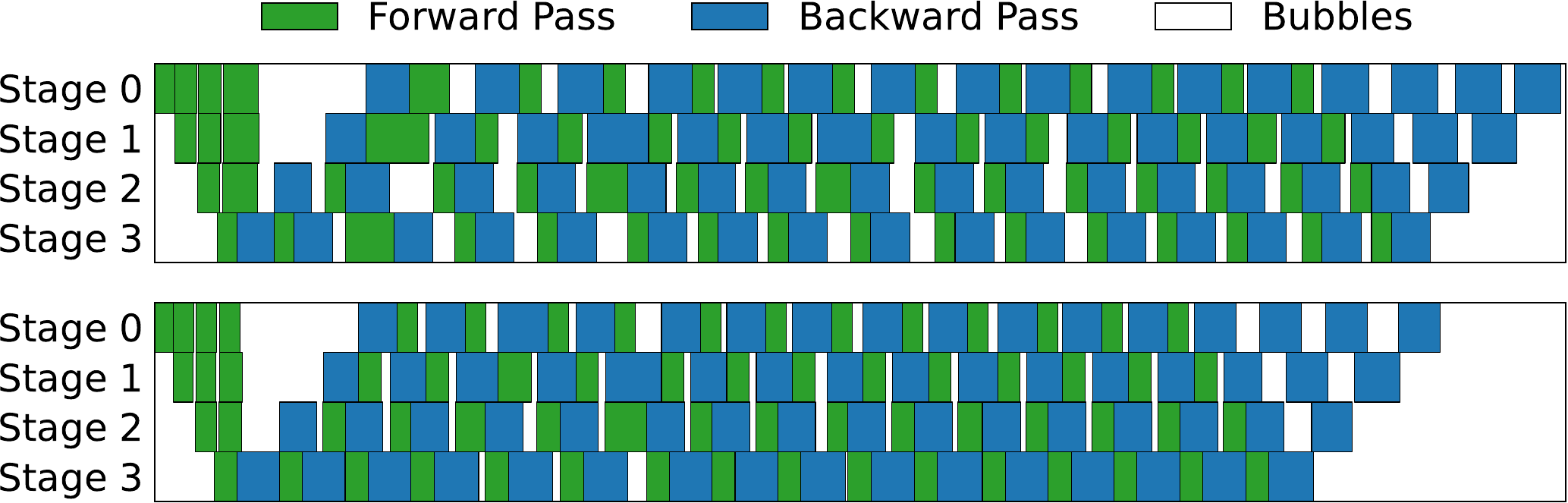}
	\vspace{-0.1in}
	\caption{\osdi{The scheduling for dp-partition (top) and \pname{}'s partition (bottom). We train  GPT-23B with 16 micro batch size.}}
	\label{fig:partition_details}
\end{figure}

\wj{
Figure~\ref{fig:partition_eval} shows the throughput comparison of  dp-partitioning and \pname{}'s partitioning.
We use GPT-13B and GPT-23B models on the NVIDIA-NVLink cluster.
}
\pname{}'s partitioning increases the throughput by 1.1$\times$--1.14$\times$, and 1.16$\times$--1.23$\times$
for the 13B and 23B models respectively.
The dp-partitioning scheme may cause uneven execution times across pipeline stages, negatively impacting overall training performance.
\wj{
Figure~\ref{fig:partition_details} shows that enabling \pname{}'s partitioning reduces the bubbles in the pipeline by 30\%.
}
Moreover, \pname{} brings more benefits for larger models because training smaller models requires less GPU memory, leading to 
lower or even no recomputation overhead, thereby 
alleviating the issue of load unbalancing across stages.


\subsubsection{Contribution of each technique}
\label{eval:breakdown}

\begin{figure}
	\centering
	\includegraphics[width=3.2in, height=1.1in]{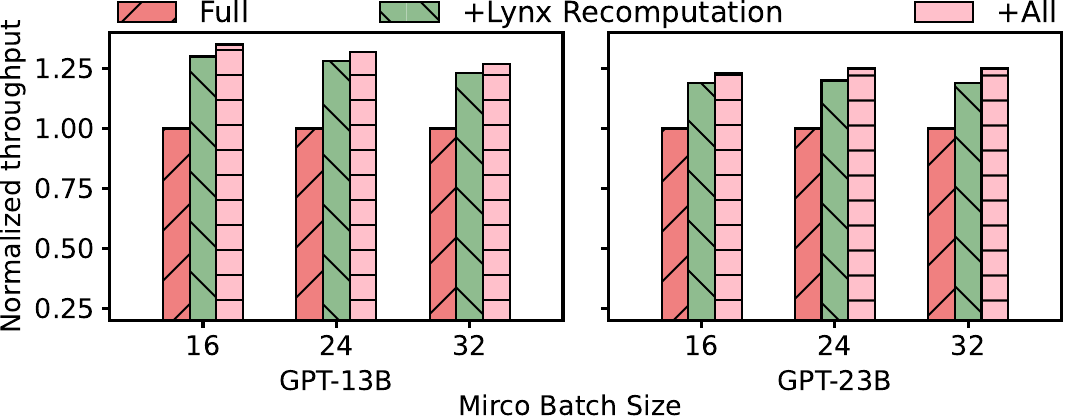}
	\caption{Performance contribution of each technique.}
	\label{fig:breakdown}
	\vspace{-0.1in}
\end{figure}

\wj{
	Figure~\ref{fig:breakdown} shows the impact of each optimization on the overall training throughput. 
	+recomputation only applies heuristic recomputation, while +All further incorporates recomputation-aware model partitioning in \pname{}.
	Specifically, +recomputation and +All enhance throughput by 1.19$\times$--1.3$\times$ and 1.25$\times$--1.35$\times$ compared to Full Recomputation, contributing an average of 77\% and 23\% to the overall performance improvement, respectively.
	This highlights the necessity of each technique.
	Moreover, model partitioning is more effective for larger models. For example, it contributes 25\% of the performance improvement on the 23B model compared to 18\% on the 13B model.
}

\subsection{The Effectiveness of \pname{}-HEU}
\label{eval:heu}
\begin{figure}
	\centering
	\includegraphics[width=3.2in]{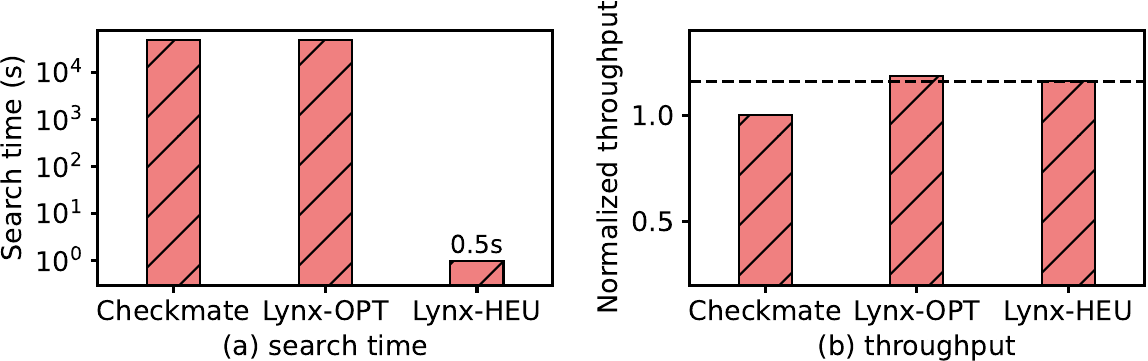}
	\caption{The effect of \pname{}-HEU on GPT-300M model.}
	\label{fig:effect_heu}
	\vspace{-0.1in}
\end{figure}

\wj{
Figure~\ref{fig:effect_heu} shows the policy search time and model training throughput of \textit{Checkmate}, \pname{} with optimal recomputation scheduling (\textit{\pname{}-OPT}), and \pname{} with heuristic recomputation scheduling (\textit{\pname{}-HEU}).
Checkmate uses MILP to determine the recomputation policy that minimizes additional recomputation costs, without accounting for overlapped recomputation (\S\ref{sec:limitations}).
We use the small GPT-300M model with micro batch sizes of 32 on the NVIDIA-NVLink cluster. 
Figure~\ref{fig:effect_heu}(a) shows that Checkmate and \textit{\pname{}-OPT} require 14 hours to find the best policy, while \textit{\pname{}-HEU} only takes 0.5 seconds.
\textit{\pname{}-HEU} achieves 97.8\% of the training throughput of  \textit{\pname{}-OPT} while reducing the search time by 99.99\%.
This demonstrates the effectiveness and practicality of \textit{\pname{}-HEU}, as it can achieve a training throughput close to the optimal in an acceptable amount of time.
}


\subsection{Sensitivity Analysis}
\label{eval:sensitivity}

\begin{figure}[t]
	\centering
	\includegraphics[width=3.2in, height=1.1in]{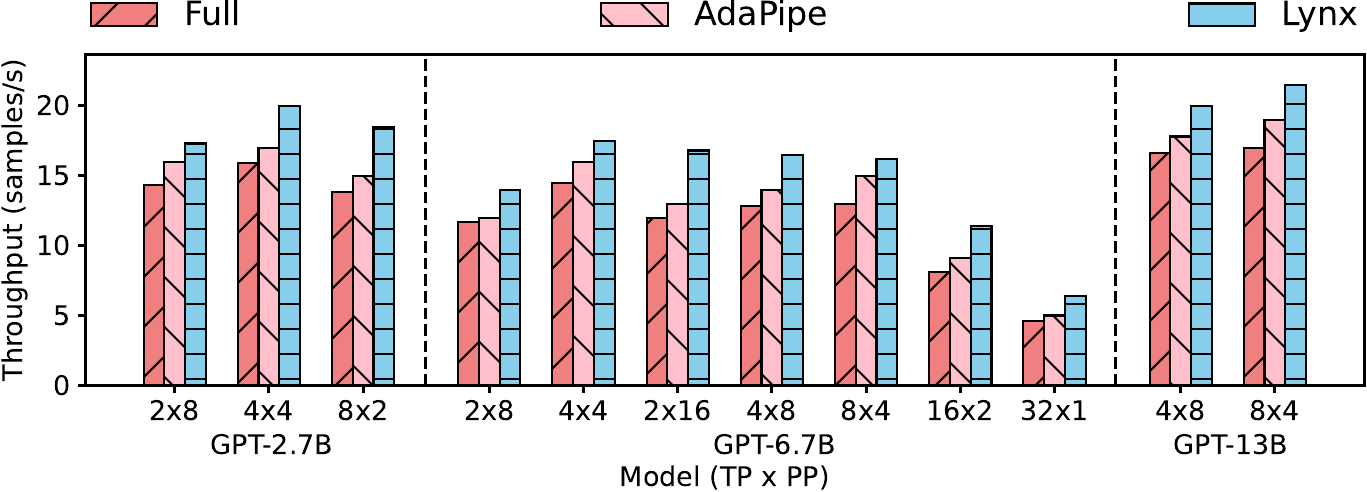}
	\vspace{-0.1in}
	\caption{Training on Ascend cluster with 8 mbs.}
	\label{fig:ascend}
	\vspace{-0.1in}
\end{figure}

\begin{figure}
	\centering
	\includegraphics[width=3.2in, height=1in]{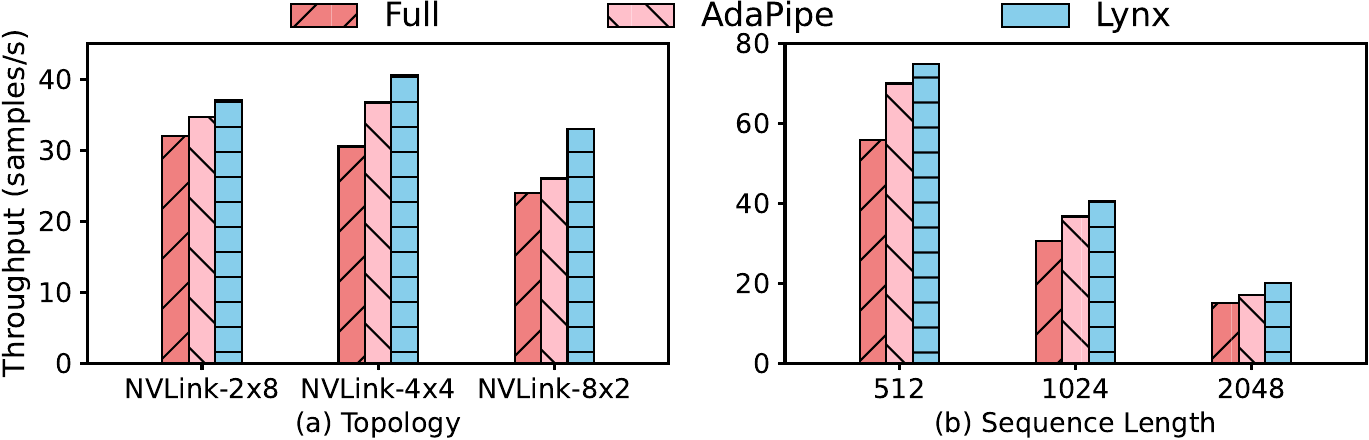}
	\vspace{-0.1in}
	\caption{Parallelism policy and sequence length on GPT-7B with micro batch size 32.}
	\label{fig:sensitivity}
	\vspace{-0.1in}
\end{figure}

\osdi{
\textbf{Accelerator types.}
Besides the NVIDIA GPUs, \pname{} is applicable to other processing units.
Figure~\ref{fig:ascend} shows that \pname{} consistently outperforms other systems on the Ascend NPU cluster. It speeds up 1.15$\times$--1.35$\times$ and 1.08$\times$--1.23$\times$ compared to Full Recomputation and AdaPipe, respectively.
Notably, \pname{} performs better on Ascend clusters than NVIDIA-NVlink devices due to higher TP communication, enabling more overlapped recomputations during communication.

\textbf{Parallelism policy.}
We configure different parallelism policy by changing the levels of tensor parallelism (TP) and pipeline parallelism (PP). 
The former equals the number of accelerators used within a single machine and the latter equals the number of machines used. 
Figure~\ref{fig:sensitivity}(a) shows that 
\pname{} outperforms other baselines by 1.06$\times$-1.37$\times$ in throughput across parallelism strategies on NVIDIA-NVLink.
Besides, under the same total number of GPUs, \pname{} shows greater speedup
with larger TP, as increased communication time allows for more effective recomputation during communication.
Figure~\ref{fig:ascend} shows similar observations for \pname{} on NPUs.

\textbf{Sequence length.}
Figure~\ref{fig:sensitivity}(b) shows that \pname{} consistently outperforms all counterparts across a range of sequence lengths from 512 to 2048.
Additionally, increasing the sequence length slows down training throughput as it raises the complexity of training.
}

\subsection{Overhead Analysis}
\label{eval:overhead}

\begin{figure}[t]
	\centering
	\includegraphics[width=3.2in]{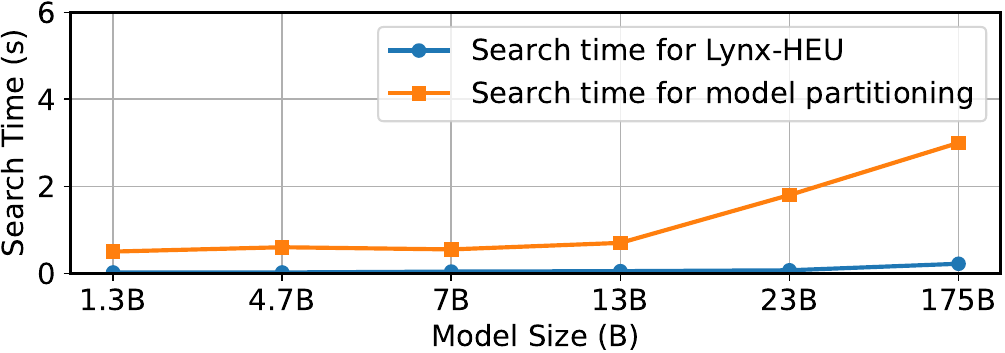}
	\vspace{-0.1in}
	\caption{Policy search time.}
	\label{fig:overhead}
	\vspace{-0.2in}
\end{figure}

\osdi{
\textbf{Profiling time.} 
\pname{} requires offline profiling of the model and collection of statistics for each operator. 
It introduces a small time overhead equivalent to several iterations of training (\S\ref{sec:overview}). 
In our experiments, profiling models ranging from 1.3B to 23B parameters takes only a few minutes due to our optimizations outlined in \S\ref{sec:implementation}. The profiling time is negligible compared to the total training time.


\textbf{Search time for \heu{}.} 
\heu{} can generate a solution within 1 seconds for models ranging from 1.3B to 175B parameters.
Moreover, Figure~\ref{fig:overhead} shows that the search time for \heu{}
remains consistent across different model sizes, demonstrating its scalability.

\textbf{Search time for model partitioning.} 
Determining the model partitioning policy, 
requires multiple invocations of \pname{}'s recomputation scheduling mechanism.
Figure~\ref{fig:overhead} shows that \heu{} takes less than 3 seconds to determine both partitioning and recomputation policies, even for large-scale models like the 175B parameter model.}

\section{Discussion}
\label{sec:discussion}
\textbf{Applicability to new techniques.} 
Other parallel techniques, like sequence parallelism (SP)~\cite{activation-recomputation-MLsys23}, are also employed in large model training. SP partitions tensors along the sequence dimension to decrease computational and memory demands for activations.
Our experiments demonstrate that \pname{} achieves an additional 10\% speedup when SP is incorporated on top of TP. This is because SP decreases the execution time of each operator, providing more opportunities for overlapping recomputation.

\textbf{Applicability to new hardwares.}
AI accelerators with extreme training performance, such as the NVIDIA GH200~\cite{gh200} and B200~\cite{b200}, are becoming available. 
Moreover, new AI training systems, such as NVIDIA DGX SuperPOD~\cite{superpod} and Google TPUv4 Pods~\cite{TPU}, have been proposed,
comprising thousands of high-performance AI accelerators. These systems may enable scaling tensor parallelism to more than eight GPUs, thereby increasing communication pressure.
In these scenarios, we believe that the techniques proposed in \pname{} will be more effective due to increased computing speed and high communication overhead.

\fast{
\textbf{Applicability to other mainstream models.}
\pname{} is applicable to most mainstream models, as they are typically designed with repeated structures (e.g., GPT series, LLaMA series, PaLM, T5, ViT).
}

\section{Related Work}

\textbf{Recomputation, swapping and compression techniques.}
Prior work uses data recomputation to extend the limited capacity of GPU memory~\cite{checkpointing-arxiv16,checkmate19,activation-recomputation-MLsys23,Deepspeed-Checkpointing,AdaPipe}.
\pname{} follows this way but can further reduce computational overhead by overlapping recomputation with communication.
Data swapping~\cite{VDNN-MICRO16, MoDNN-DATE18, CSwap,Swapadvisor-ASPLOS20, FlashNeuron-FAST21, Behemoth-FAST21} and their combination with recomputation~\cite{SuperNeurons-PPoPP18,Layup,Capuchin-ASPLOS20,AutoTM-ASPLOS20,HOME-TC23} can be also leveraged to  minimize GPU memory footprint.
These techniques complement to our approach.
Compression techniques are  widely used  to eliminate data redundancy during DNN training~\cite{CSwap,CSWAP+-TPDS22,GIST-ISCA18,yang2021auto}, but they may compromise model accuracy. 

\textbf{Data parallelism, tensor parallelism, and pipeline parallelism.}
DP partitions input samples among different workers~\cite{PyTorch,Tensorflow-OSDI16,zico,activation-recomputation-MLsys23}.  
However, as the size of the model grows, these approaches will suffer from  communication bottlenecks~\cite{MPress-HPCA23,Merak-TPDS23}. 
TP splits model weight matrices and assign them to different devices~\cite{Whale-ATC22,PTD-P-SC21,Megatron-Arxiv19,Gpipe-NIPS19,DAPPLE-PPoPP21,Pipedream}.
PP partitions  a model into sub-modules to multiple GPUs and transfer the output of each module to the next device~\cite{Gpipe-NIPS19,DAPPLE-PPoPP21,Megatron-Arxiv19,Chimera-SC21,Pipedream,Pipedram-2-PMLR21}.
Existing works also consider evenly partitioning models to achieve the computation balance ~\cite{Pipedream,PTD-P-SC21,pp-nips20,deepspeed-Github}.
\fast{However, Lynx considers the impact of recomputation on performance when partitioning the model into different stages, whereas other approaches do not.}

\textbf{Overlapping computation within communication.}
Previous studies apply a variety of loop analysis and transformation techniques to extract loops containing only independent communication and computation for overlapping~\cite{pipe-1,pipe-2}.
Some works accelerate DNN training through hardware~\cite{ACE-isca2021} or compiler optimizations~\cite{pipe-3}.
They are orthogonal to \pname{} as they do not consider overlapping recomputation.


\section{Conclustion}

In this paper, we propose the \pname{} framework
for large DNN model training with recomputation.
First, it reduces recomputation overhead by overlapping
recomputation with communication, which is required in 
tensor and pipeline parallelism. 
Second, we model the recomputation
scheduling problem and solve it \wj{using an integer 
linear program to achieve a near-optimal solution} based on the
heuristics that large models have identical structures
to reduce the size of solution space.
Finally, we design a model partitioning algorithm
to achieve load balancing among pipeline stages.
We evaluate the performance of \pname{} across
different models using both
NVLink and PCIe connected GPU clusters. The results
show that \pname{} outperforms the existing
approaches by
up to 1.37$\times$.

	\bibliography{refs}


\begin{thebibliography}{68}


\ifx \showCODEN    \undefined \def \showCODEN     #1{\unskip}     \fi
\ifx \showDOI      \undefined \def \showDOI       #1{#1}\fi
\ifx \showISBNx    \undefined \def \showISBNx     #1{\unskip}     \fi
\ifx \showISBNxiii \undefined \def \showISBNxiii  #1{\unskip}     \fi
\ifx \showISSN     \undefined \def \showISSN      #1{\unskip}     \fi
\ifx \showLCCN     \undefined \def \showLCCN      #1{\unskip}     \fi
\ifx \shownote     \undefined \def \shownote      #1{#1}          \fi
\ifx \showarticletitle \undefined \def \showarticletitle #1{#1}   \fi
\ifx \showURL      \undefined \def \showURL       {\relax}        \fi
\providecommand\bibfield[2]{#2}
\providecommand\bibinfo[2]{#2}
\providecommand\natexlab[1]{#1}
\providecommand\showeprint[2][]{arXiv:#2}

\bibitem[Bae et~al\mbox{.}(2021)]%
        {FlashNeuron-FAST21}
\bibfield{author}{\bibinfo{person}{Jonghyun Bae}, \bibinfo{person}{Jongsung
  Lee}, \bibinfo{person}{Yunho Jin}, \bibinfo{person}{Sam Son},
  \bibinfo{person}{Shine Kim}, \bibinfo{person}{Hakbeom Jang},
  \bibinfo{person}{Tae~Jun Ham}, {and} \bibinfo{person}{Jae~W Lee}.}
  \bibinfo{year}{2021}\natexlab{}.
\newblock \showarticletitle{{Flash ``Neuron:SSD-Enabled Large-Batch Training of
  Very Deep Neural Networks}}. In \bibinfo{booktitle}{\emph{Proceedings of the
  19th USENIX Conference on File and Storage Technologies}}.
\newblock


\bibitem[Brown(2020)]%
        {gpt3}
\bibfield{author}{\bibinfo{person}{Tom~B Brown}.}
  \bibinfo{year}{2020}\natexlab{}.
\newblock \showarticletitle{{Language Models are Few-shot Learners}}.
\newblock \bibinfo{journal}{\emph{arXiv preprint arXiv:2005.14165}}
  (\bibinfo{year}{2020}).
\newblock


\bibitem[Chen et~al\mbox{.}(2020)]%
        {gptimage-icml20}
\bibfield{author}{\bibinfo{person}{Mark Chen}, \bibinfo{person}{Alec Radford},
  \bibinfo{person}{Rewon Child}, \bibinfo{person}{Jeffrey Wu},
  \bibinfo{person}{Heewoo Jun}, \bibinfo{person}{David Luan}, {and}
  \bibinfo{person}{Ilya Sutskever}.} \bibinfo{year}{2020}\natexlab{}.
\newblock \showarticletitle{{Generative Pretraining from Pixels}}. In
  \bibinfo{booktitle}{\emph{Proceedings of the International conference on
  machine learning}}.
\newblock


\bibitem[Chen et~al\mbox{.}(2022)]%
        {CSWAP+-TPDS22}
\bibfield{author}{\bibinfo{person}{Ping Chen}, \bibinfo{person}{Shuibing He},
  \bibinfo{person}{Xuechen Zhang}, \bibinfo{person}{Shuaiben Chen},
  \bibinfo{person}{Peiyi Hong}, \bibinfo{person}{Yanlong Yin}, {and}
  \bibinfo{person}{Xian-He Sun}.} \bibinfo{year}{2022}\natexlab{}.
\newblock \showarticletitle{{Accelerating Tensor Swapping in GPUs With
  Self-Tuning Compression}}.
\newblock \bibinfo{journal}{\emph{IEEE Transactions on Parallel and Distributed
  Systems}} (\bibinfo{year}{2022}).
\newblock


\bibitem[Chen et~al\mbox{.}(2021)]%
        {CSwap}
\bibfield{author}{\bibinfo{person}{Ping Chen}, \bibinfo{person}{Shuibing He},
  \bibinfo{person}{Xuechen Zhang}, \bibinfo{person}{Shuaiben Chen},
  \bibinfo{person}{Peiyi Hong}, \bibinfo{person}{Yanlong Yin},
  \bibinfo{person}{Xian-He Sun}, {and} \bibinfo{person}{Gang Chen}.}
  \bibinfo{year}{2021}\natexlab{}.
\newblock \showarticletitle{{CSWAP: A Self-Tuning Compression Framework for
  Accelerating Tensor Swapping in GPUs}}. In
  \bibinfo{booktitle}{\emph{Proceedings of the 2021 IEEE International
  Conference on Cluster Computing}}.
\newblock


\bibitem[Chen et~al\mbox{.}(2016)]%
        {checkpointing-arxiv16}
\bibfield{author}{\bibinfo{person}{Tianqi Chen}, \bibinfo{person}{Bing Xu},
  \bibinfo{person}{Chiyuan Zhang}, {and} \bibinfo{person}{Carlos Guestrin}.}
  \bibinfo{year}{2016}\natexlab{}.
\newblock \showarticletitle{{Training Deep Nets with Sublinear Memory Cost}}.
\newblock \bibinfo{journal}{\emph{arXiv preprint arXiv:1604.06174}}
  (\bibinfo{year}{2016}).
\newblock


\bibitem[Chen et~al\mbox{.}(2023)]%
        {icache}
\bibfield{author}{\bibinfo{person}{Weijian Chen}, \bibinfo{person}{Shuibing
  He}, \bibinfo{person}{Yaowen Xu}, \bibinfo{person}{Xuechen Zhang},
  \bibinfo{person}{Siling Yang}, \bibinfo{person}{Shuang Hu},
  \bibinfo{person}{Xian-He Sun}, {and} \bibinfo{person}{Gang Chen}.}
  \bibinfo{year}{2023}\natexlab{}.
\newblock \showarticletitle{{iCache: An Importance-Sampling-Informed Cache for
  Accelerating I/O-Bound DNN Model Training}}. In
  \bibinfo{booktitle}{\emph{Proceedings of the 2023 IEEE International
  Symposium on High-Performance Computer Architecture}}.
\newblock


\bibitem[Chen et~al\mbox{.}(2018)]%
        {MoDNN-DATE18}
\bibfield{author}{\bibinfo{person}{Xiaoming Chen}, \bibinfo{person}{Danny~Z.
  Chen}, {and} \bibinfo{person}{Xiaobo~Sharon Hu}.}
  \bibinfo{year}{2018}\natexlab{}.
\newblock \showarticletitle{{MoDNN: Memory Optimal DNN Training on GPUs}}. In
  \bibinfo{booktitle}{\emph{Proceedings of the Design, Automation and Test in
  Europe Conference and Exhibition}}.
\newblock


\bibitem[Chowdhery et~al\mbox{.}(2023)]%
        {palm}
\bibfield{author}{\bibinfo{person}{Aakanksha Chowdhery},
  \bibinfo{person}{Sharan Narang}, \bibinfo{person}{Jacob Devlin},
  \bibinfo{person}{Maarten Bosma}, \bibinfo{person}{Gaurav Mishra},
  \bibinfo{person}{Adam Roberts}, \bibinfo{person}{Paul Barham},
  \bibinfo{person}{Hyung~Won Chung}, \bibinfo{person}{Charles Sutton},
  \bibinfo{person}{Sebastian Gehrmann}, {et~al\mbox{.}}}
  \bibinfo{year}{2023}\natexlab{}.
\newblock \showarticletitle{{Palm: Scaling Language Modeling with Pathways}}.
\newblock \bibinfo{journal}{\emph{Journal of Machine Learning Research}}
  (\bibinfo{year}{2023}).
\newblock


\bibitem[ColossalAI(2024)]%
        {colossalai}
\bibfield{author}{\bibinfo{person}{ColossalAI}.}
  \bibinfo{year}{2024}\natexlab{}.
\newblock \bibinfo{title}{{ColossalAI}}.
\newblock \bibinfo{howpublished}{\url{https://colossalai.org/}}.
\newblock
\urldef\tempurl%
\url{https://colossalai.org/}
\showURL{%
\tempurl}


\bibitem[Danalis et~al\mbox{.}(2005)]%
        {pipe-1}
\bibfield{author}{\bibinfo{person}{A. Danalis}, \bibinfo{person}{K.-Y. Kim},
  \bibinfo{person}{L. Pollock}, {and} \bibinfo{person}{M. Swany}.}
  \bibinfo{year}{2005}\natexlab{}.
\newblock \showarticletitle{{Transformations to Parallel Codes for
  Communication-Computation Overlap}}. In \bibinfo{booktitle}{\emph{Proceedings
  of the 2005 ACM/IEEE Conference on Supercomputing}}.
\newblock


\bibitem[Deepspeed(2023)]%
        {Deepspeed-Checkpointing}
\bibfield{author}{\bibinfo{person}{Deepspeed}.}
  \bibinfo{year}{2023}\natexlab{}.
\newblock \bibinfo{title}{Activation Checkpointing}.
\newblock
  \bibinfo{howpublished}{\url{https://deepspeed.readthedocs.io/en/stable/activation-checkpointing.html}}.
\newblock
\urldef\tempurl%
\url{https://deepspeed.readthedocs.io/en/stable/activation-checkpointing.html}
\showURL{%
\tempurl}


\bibitem[Deepspeed-Megatron(2024)]%
        {Deepspeed-Partition}
\bibfield{author}{\bibinfo{person}{Deepspeed-Megatron}.}
  \bibinfo{year}{2024}\natexlab{}.
\newblock \bibinfo{title}{Pipeline Parallelism}.
\newblock
  \bibinfo{howpublished}{\url{https://www.deepspeed.ai/tutorials/pipeline/}}.
\newblock
\urldef\tempurl%
\url{https://www.deepspeed.ai/tutorials/pipeline/}
\showURL{%
\tempurl}


\bibitem[Fan et~al\mbox{.}(2021)]%
        {DAPPLE-PPoPP21}
\bibfield{author}{\bibinfo{person}{Shiqing Fan}, \bibinfo{person}{Yi Rong},
  \bibinfo{person}{Chen Meng}, \bibinfo{person}{Zongyan Cao},
  \bibinfo{person}{Siyu Wang}, \bibinfo{person}{Zhen Zheng},
  \bibinfo{person}{Chuan Wu}, \bibinfo{person}{Guoping Long},
  \bibinfo{person}{Jun Yang}, \bibinfo{person}{Lixue Xia}, {et~al\mbox{.}}}
  \bibinfo{year}{2021}\natexlab{}.
\newblock \showarticletitle{{DAPPLE: A Pipelined Data Parallel Approach for
  Training Large Models}}. In \bibinfo{booktitle}{\emph{Proceedings of the 26th
  ACM SIGPLAN Symposium on Principles and Practice of Parallel Programming}}.
\newblock


\bibitem[Google(2024)]%
        {TPU}
\bibfield{author}{\bibinfo{person}{Google}.} \bibinfo{year}{2024}\natexlab{}.
\newblock \bibinfo{title}{{Google showcases Cloud TPU v4 Pods for large model
  training}}.
\newblock
  \bibinfo{howpublished}{\url{https://cloud.google.com/blog/topics/tpus/google-showcases-cloud-/tpu-v4-pods-for-large-model-training}}.
\newblock
\urldef\tempurl%
\url{https://cloud.google.com/blog/topics/tpus/google-showcases-cloud-tpu-v4-pods-for-large-model-training}
\showURL{%
\tempurl}


\bibitem[Guo et~al\mbox{.}(2016)]%
        {pipe-2}
\bibfield{author}{\bibinfo{person}{J. Guo}, \bibinfo{person}{Q. Yi},
  \bibinfo{person}{J. Meng}, \bibinfo{person}{J. Zhang}, {and}
  \bibinfo{person}{P. Balaji}.} \bibinfo{year}{2016}\natexlab{}.
\newblock \showarticletitle{{Compiler-Assisted Overlapping of Communication and
  Computation in MPI Applications}}. In \bibinfo{booktitle}{\emph{Proceedings
  of the 2016 IEEE International Conference on Cluster Computing}}.
\newblock


\bibitem[gurobi(2024)]%
        {gurobi}
\bibfield{author}{\bibinfo{person}{gurobi}.} \bibinfo{year}{2024}\natexlab{}.
\newblock \bibinfo{title}{Gurobi}.
\newblock \bibinfo{howpublished}{\url{https://www.gurobi.com/}}.
\newblock


\bibitem[He et~al\mbox{.}(2023)]%
        {HOME-TC23}
\bibfield{author}{\bibinfo{person}{Shuibing He}, \bibinfo{person}{Ping Chen},
  \bibinfo{person}{Shuaiben Chen}, \bibinfo{person}{Zheng Li},
  \bibinfo{person}{Siling Yang}, \bibinfo{person}{Weijian Chen}, {and}
  \bibinfo{person}{Lidan Shou}.} \bibinfo{year}{2023}\natexlab{}.
\newblock \showarticletitle{{HOME: A Holistic GPU Memory Management Framework
  for Deep Learning}}.
\newblock \bibinfo{journal}{\emph{IEEE Trans. Comput.}} (\bibinfo{year}{2023}).
\newblock


\bibitem[Hildebrand et~al\mbox{.}(2020)]%
        {AutoTM-ASPLOS20}
\bibfield{author}{\bibinfo{person}{Mark Hildebrand}, \bibinfo{person}{Jawad
  Khan}, \bibinfo{person}{Sanjeev Trika}, \bibinfo{person}{Jason Lowe-Power},
  {and} \bibinfo{person}{Venkatesh Akella}.} \bibinfo{year}{2020}\natexlab{}.
\newblock \showarticletitle{{AutOTM: Automatic Tensor Movement in Heterogeneous
  Memory Systems Using Integer Linear Programming}}. In
  \bibinfo{booktitle}{\emph{Proceedings of the International Conference on
  Architectural Support for Programming Languages and Operating Systems}}.
\newblock


\bibitem[Huang et~al\mbox{.}(2020)]%
        {Swapadvisor-ASPLOS20}
\bibfield{author}{\bibinfo{person}{Chien~Chin Huang}, \bibinfo{person}{Gu Jin},
  {and} \bibinfo{person}{Jinyang Li}.} \bibinfo{year}{2020}\natexlab{}.
\newblock \showarticletitle{{SwapAdvisor: Push Deep Learning Beyond the GPU
  Memory Limit via Smart Swapping}}. In \bibinfo{booktitle}{\emph{Proceedings
  of the International Conference on Architectural Support for Programming
  Languages and Operating Systems}}.
\newblock


\bibitem[Huang et~al\mbox{.}(2019)]%
        {Gpipe-NIPS19}
\bibfield{author}{\bibinfo{person}{Yanping Huang}, \bibinfo{person}{Youlong
  Cheng}, \bibinfo{person}{Ankur Bapna}, \bibinfo{person}{Orhan Firat},
  \bibinfo{person}{Dehao Chen}, \bibinfo{person}{Mia Chen},
  \bibinfo{person}{HyoukJoong Lee}, \bibinfo{person}{Jiquan Ngiam},
  \bibinfo{person}{Quoc~V Le}, \bibinfo{person}{Yonghui Wu}, {et~al\mbox{.}}}
  \bibinfo{year}{2019}\natexlab{}.
\newblock \showarticletitle{{Gpipe: Efficient Training of Giant Neural Networks
  Using Pipeline Parallelism}}. In \bibinfo{booktitle}{\emph{Proceedings of the
  Advances in neural information processing systems}}.
\newblock


\bibitem[Huawei(2024a)]%
        {MindSpeed}
\bibfield{author}{\bibinfo{person}{Huawei}.} \bibinfo{year}{2024}\natexlab{a}.
\newblock \bibinfo{title}{Ascend MindSpeed-LLM}.
\newblock \bibinfo{howpublished}{\url{https://gitee.com/ascend/MindSpeed-LLM}}.
\newblock
\urldef\tempurl%
\url{https://gitee.com/ascend/MindSpeed-LLM}
\showURL{%
\tempurl}


\bibitem[Huawei(2024b)]%
        {MindSpore}
\bibfield{author}{\bibinfo{person}{Huawei}.} \bibinfo{year}{2024}\natexlab{b}.
\newblock \bibinfo{title}{{MindSpore}}.
\newblock \bibinfo{howpublished}{\url{https://github.com/mindspore-ai}}.
\newblock
\urldef\tempurl%
\url{https://github.com/mindspore-ai}
\showURL{%
\tempurl}


\bibitem[Jain et~al\mbox{.}(2018)]%
        {GIST-ISCA18}
\bibfield{author}{\bibinfo{person}{Animesh Jain}, \bibinfo{person}{Amar
  Phanishayee}, \bibinfo{person}{Jason Mars}, \bibinfo{person}{Lingjia Tang},
  {and} \bibinfo{person}{Gennady Pekhimenko}.} \bibinfo{year}{2018}\natexlab{}.
\newblock \showarticletitle{{GIST: Efficient Data Encoding for Deep Neural
  Network Training}}. In \bibinfo{booktitle}{\emph{Proceedings of the
  International Symposium on Computer Architecture}}.
\newblock


\bibitem[Jain et~al\mbox{.}(2019)]%
        {checkmate19}
\bibfield{author}{\bibinfo{person}{Paras Jain}, \bibinfo{person}{Ajay Jain},
  \bibinfo{person}{Aniruddha Nrusimha}, \bibinfo{person}{Amir Gholami},
  \bibinfo{person}{Pieter Abbeel}, \bibinfo{person}{Kurt Keutzer},
  \bibinfo{person}{Ion Stoica}, {and} \bibinfo{person}{Joseph~E. Gonzalez}.}
  \bibinfo{year}{2019}\natexlab{}.
\newblock \showarticletitle{{Checkmate: Breaking the Memory Wall with Optimal
  Tensor Rematerialization}}.
\newblock \bibinfo{journal}{\emph{arXiv preprint arXiv:1910.02653}}
  (\bibinfo{year}{2019}).
\newblock


\bibitem[Jia et~al\mbox{.}(2022)]%
        {Whale-ATC22}
\bibfield{author}{\bibinfo{person}{Xianyan Jia}, \bibinfo{person}{Le Jiang},
  \bibinfo{person}{Ang Wang}, \bibinfo{person}{Wencong Xiao},
  \bibinfo{person}{Ziji Shi}, \bibinfo{person}{Jie Zhang},
  \bibinfo{person}{Xinyuan Li}, \bibinfo{person}{Langshi Chen},
  \bibinfo{person}{Yong Li}, \bibinfo{person}{Zhen Zheng}, {et~al\mbox{.}}}
  \bibinfo{year}{2022}\natexlab{}.
\newblock \showarticletitle{{Whale: Efficient Giant Model Training over
  Heterogeneous GPUs}}. In \bibinfo{booktitle}{\emph{Proceedings of the 2022
  USENIX Annual Technical Conference}}.
\newblock


\bibitem[Jiang et~al\mbox{.}(2019)]%
        {Layup}
\bibfield{author}{\bibinfo{person}{Wenbin Jiang}, \bibinfo{person}{Yang Ma},
  \bibinfo{person}{Bo Liu}, \bibinfo{person}{Haikun Liu},
  \bibinfo{person}{Bing~Bing Zhou}, \bibinfo{person}{Jian Zhu},
  \bibinfo{person}{Song Wu}, {and} \bibinfo{person}{Hai Jin}.}
  \bibinfo{year}{2019}\natexlab{}.
\newblock \showarticletitle{{Layup: Layer-adaptive and Multi-type
  Intermediate-oriented Memory Optimization for GPU-based CNNs}}.
\newblock \bibinfo{journal}{\emph{ACM Transactions on Architecture and Code
  Optimization}} (\bibinfo{year}{2019}).
\newblock


\bibitem[Jiang et~al\mbox{.}(2024)]%
        {MegaScale-arxiv24}
\bibfield{author}{\bibinfo{person}{Ziheng Jiang}, \bibinfo{person}{Haibin Lin},
  \bibinfo{person}{Yinmin Zhong}, \bibinfo{person}{Qi Huang},
  \bibinfo{person}{Yangrui Chen}, \bibinfo{person}{Zhi Zhang},
  \bibinfo{person}{Yanghua Peng}, \bibinfo{person}{Xiang Li},
  \bibinfo{person}{Cong Xie}, \bibinfo{person}{Shibiao Nong}, {et~al\mbox{.}}}
  \bibinfo{year}{2024}\natexlab{}.
\newblock \showarticletitle{{MegaScale: Scaling Large Language Model Training
  to More Than 10,000 GPUs}}.
\newblock \bibinfo{journal}{\emph{arXiv preprint arXiv:2402.15627}}
  (\bibinfo{year}{2024}).
\newblock


\bibitem[Kaplan et~al\mbox{.}(2020)]%
        {scaling}
\bibfield{author}{\bibinfo{person}{Jared Kaplan}, \bibinfo{person}{Sam
  McCandlish}, \bibinfo{person}{Tom Henighan}, \bibinfo{person}{Tom~B Brown},
  \bibinfo{person}{Benjamin Chess}, \bibinfo{person}{Rewon Child},
  \bibinfo{person}{Scott Gray}, \bibinfo{person}{Alec Radford},
  \bibinfo{person}{Jeffrey Wu}, {and} \bibinfo{person}{Dario Amodei}.}
  \bibinfo{year}{2020}\natexlab{}.
\newblock \showarticletitle{{Scaling Laws for Neural Language Models}}.
\newblock \bibinfo{journal}{\emph{arXiv preprint arXiv:2001.08361}}
  (\bibinfo{year}{2020}).
\newblock


\bibitem[Kim et~al\mbox{.}(2021)]%
        {Behemoth-FAST21}
\bibfield{author}{\bibinfo{person}{Shine Kim}, \bibinfo{person}{Yunho Jin},
  \bibinfo{person}{Gina Sohn}, \bibinfo{person}{Jonghyun Bae},
  \bibinfo{person}{Tae~Jun Ham}, {and} \bibinfo{person}{Jae~W Lee}.}
  \bibinfo{year}{2021}\natexlab{}.
\newblock \showarticletitle{{Behemoth: a Flash-centric Training Accelerator for
  Extreme-scale DNNs}}. In \bibinfo{booktitle}{\emph{Proceedings of the 19th
  USENIX Conference on File and Storage Technologies}}.
\newblock


\bibitem[Kingma and Ba(2014)]%
        {Adam}
\bibfield{author}{\bibinfo{person}{Diederik~P Kingma} {and}
  \bibinfo{person}{Jimmy Ba}.} \bibinfo{year}{2014}\natexlab{}.
\newblock \showarticletitle{{Adam: A Method for Stochastic Optimization}}.
\newblock \bibinfo{journal}{\emph{arXiv preprint arXiv:1412.6980}}
  (\bibinfo{year}{2014}).
\newblock


\bibitem[Korthikanti et~al\mbox{.}(2023)]%
        {activation-recomputation-MLsys23}
\bibfield{author}{\bibinfo{person}{Vijay~Anand Korthikanti},
  \bibinfo{person}{Jared Casper}, \bibinfo{person}{Sangkug Lym},
  \bibinfo{person}{Lawrence McAfee}, \bibinfo{person}{Michael Andersch},
  \bibinfo{person}{Mohammad Shoeybi}, {and} \bibinfo{person}{Bryan Catanzaro}.}
  \bibinfo{year}{2023}\natexlab{}.
\newblock \showarticletitle{{Reducing Activation Recomputation in Large
  Transformer Models}}. In \bibinfo{booktitle}{\emph{Proceedings of Machine
  Learning and Systems}}.
\newblock


\bibitem[Lai et~al\mbox{.}(2023)]%
        {Merak-TPDS23}
\bibfield{author}{\bibinfo{person}{Zhiquan Lai}, \bibinfo{person}{Shengwei Li},
  \bibinfo{person}{Xudong Tang}, \bibinfo{person}{Keshi Ge},
  \bibinfo{person}{Weijie Liu}, \bibinfo{person}{Yabo Duan},
  \bibinfo{person}{Linbo Qiao}, {and} \bibinfo{person}{Dongsheng Li}.}
  \bibinfo{year}{2023}\natexlab{}.
\newblock \showarticletitle{{Merak: An Efficient Distributed DNN Training
  Framework with Automated 3D Parallelism for Giant Foundation Models}}.
\newblock \bibinfo{journal}{\emph{IEEE Transactions on Parallel and Distributed
  Systems}} (\bibinfo{year}{2023}).
\newblock


\bibitem[Lambda(2020)]%
        {GPT-3}
\bibfield{author}{\bibinfo{person}{Lambda}.} \bibinfo{year}{2020}\natexlab{}.
\newblock \bibinfo{title}{{OpenAI's GPT-3 Language Model: A Technical
  Overview}}.
\newblock
  \bibinfo{howpublished}{\url{https://lambdalabs.com/blog/demystifying-gpt-3}}.
\newblock
\urldef\tempurl%
\url{https://lambdalabs.com/blog/demystifying-gpt-3}
\showURL{%
\tempurl}


\bibitem[Li and Hoefler(2021)]%
        {Chimera-SC21}
\bibfield{author}{\bibinfo{person}{Shigang Li} {and} \bibinfo{person}{Torsten
  Hoefler}.} \bibinfo{year}{2021}\natexlab{}.
\newblock \showarticletitle{{Chimera: Efficiently Training Large-Scale Neural
  Networks with Bidirectional Pipelines}}. In
  \bibinfo{booktitle}{\emph{Proceedings of the International Conference for
  High Performance Computing, Networking, Storage and Analysis}}.
\newblock


\bibitem[Liang et~al\mbox{.}(2023)]%
        {survey-TPDS23}
\bibfield{author}{\bibinfo{person}{Peng Liang}, \bibinfo{person}{Yu Tang},
  \bibinfo{person}{Xiaoda Zhang}, \bibinfo{person}{Youhui Bai},
  \bibinfo{person}{Teng Su}, \bibinfo{person}{Zhiquan Lai},
  \bibinfo{person}{Linbo Qiao}, {and} \bibinfo{person}{Dongsheng Li}.}
  \bibinfo{year}{2023}\natexlab{}.
\newblock \showarticletitle{{A Survey on Auto-Parallelism of Large-Scale Deep
  Learning Training}}.
\newblock \bibinfo{journal}{\emph{IEEE Transactions on Parallel and Distributed
  Systems}} (\bibinfo{year}{2023}).
\newblock


\bibitem[Lim et~al\mbox{.}(2021)]%
        {zico}
\bibfield{author}{\bibinfo{person}{Gangmuk Lim}, \bibinfo{person}{Jeongseob
  Ahn}, \bibinfo{person}{Wencong Xiao}, \bibinfo{person}{Youngjin Kwon}, {and}
  \bibinfo{person}{Myeongjae Jeon}.} \bibinfo{year}{2021}\natexlab{}.
\newblock \showarticletitle{{Zico: Efficient GPU Memory Sharing for Concurrent
  DNN Training}}. In \bibinfo{booktitle}{\emph{Proceedings of the 2021 USENIX
  Annual Technical Conference}}.
\newblock


\bibitem[{Mart{\'{i}}n Abadi, Paul Barham, Jianmin Chen, Zhifeng Chen, Andy
  Davis, Jeffrey Dean, Matthieu Devin, Sanjay Ghemawat, Geoffrey Irving,
  Michael Isard, Manjunath Kudlur, Josh Levenberg, Rajat Monga, Sherry Moore,
  Derek G. Murray, Benoit Steiner, Paul Tucker, Vi}(2016)]%
        {Tensorflow-OSDI16}
\bibfield{author}{\bibinfo{person}{Google~Brain {Mart{\'{i}}n Abadi, Paul
  Barham, Jianmin Chen, Zhifeng Chen, Andy Davis, Jeffrey Dean, Matthieu Devin,
  Sanjay Ghemawat, Geoffrey Irving, Michael Isard, Manjunath Kudlur, Josh
  Levenberg, Rajat Monga, Sherry Moore, Derek G. Murray, Benoit Steiner, Paul
  Tucker, Vi}}.} \bibinfo{year}{2016}\natexlab{}.
\newblock \showarticletitle{{TensorFlow: A System for Large-scale Machine
  Learning}}. In \bibinfo{booktitle}{\emph{Proceedings of the USENIX Symposium
  on Operating Systems Design and Implementation}}.
\newblock


\bibitem[Megatron-WikiText(2024)]%
        {megatron-wiki}
\bibfield{author}{\bibinfo{person}{Megatron-WikiText}.}
  \bibinfo{year}{2024}\natexlab{}.
\newblock \bibinfo{title}{{Collecting Wikipedia Training Data}}.
\newblock
\newblock
\urldef\tempurl%
\url{https://github.com/NVIDIA/Megatron-LM?tab=readme-ov-file#collecting-wikipedia-training-data}
\showURL{%
\tempurl}


\bibitem[Microsoft(2023)]%
        {deepspeed-Github}
\bibfield{author}{\bibinfo{person}{Microsoft}.}
  \bibinfo{year}{2023}\natexlab{}.
\newblock \bibinfo{title}{Megatron-DeepSpeed}.
\newblock
  \bibinfo{howpublished}{\url{https://github.com/microsoft/Megatron-DeepSpeed/tree/main}}.
\newblock
\urldef\tempurl%
\url{https://github.com/microsoft/Megatron-DeepSpeed/tree/main}
\showURL{%
\tempurl}


\bibitem[Narayanan et~al\mbox{.}(2019)]%
        {Pipedream}
\bibfield{author}{\bibinfo{person}{Deepak Narayanan}, \bibinfo{person}{Aaron
  Harlap}, \bibinfo{person}{Amar Phanishayee}, \bibinfo{person}{Vivek
  Seshadri}, \bibinfo{person}{Nikhil~R. Devanur}, \bibinfo{person}{Gregory~R.
  Ganger}, \bibinfo{person}{Phillip~B. Gibbons}, {and} \bibinfo{person}{Matei
  Zaharia}.} \bibinfo{year}{2019}\natexlab{}.
\newblock \showarticletitle{{PipeDream: Generalized Pipeline Parallelism for
  DNN Training}}. In \bibinfo{booktitle}{\emph{Proceedings of the 27th ACM
  Symposium on Operating Systems Principles}}.
\newblock


\bibitem[Narayanan et~al\mbox{.}(2021a)]%
        {Pipedram-2-PMLR21}
\bibfield{author}{\bibinfo{person}{Deepak Narayanan}, \bibinfo{person}{Amar
  Phanishayee}, \bibinfo{person}{Kaiyu Shi}, \bibinfo{person}{Xie Chen}, {and}
  \bibinfo{person}{Matei Zaharia}.} \bibinfo{year}{2021}\natexlab{a}.
\newblock \showarticletitle{{Memory-efficient Pipeline-parallel DNN Training}}.
  In \bibinfo{booktitle}{\emph{Proceedings of the International Conference on
  Machine Learning}}. PMLR.
\newblock


\bibitem[Narayanan et~al\mbox{.}(2021b)]%
        {PTD-P-SC21}
\bibfield{author}{\bibinfo{person}{Deepak Narayanan}, \bibinfo{person}{Mohammad
  Shoeybi}, \bibinfo{person}{Jared Casper}, \bibinfo{person}{Patrick
  LeGresley}, \bibinfo{person}{Mostofa Patwary}, \bibinfo{person}{Vijay
  Korthikanti}, \bibinfo{person}{Dmitri Vainbrand}, \bibinfo{person}{Prethvi
  Kashinkunti}, \bibinfo{person}{Julie Bernauer}, \bibinfo{person}{Bryan
  Catanzaro}, {et~al\mbox{.}}} \bibinfo{year}{2021}\natexlab{b}.
\newblock \showarticletitle{{Efficient Large-scale Language Model Training on
  Gpu Clusters Using Megatron-LM}}. In \bibinfo{booktitle}{\emph{Proceedings of
  the International Conference for High Performance Computing, Networking,
  Storage and Analysis}}.
\newblock


\bibitem[NVIDIA(2023)]%
        {gh200}
\bibfield{author}{\bibinfo{person}{NVIDIA}.} \bibinfo{year}{2023}\natexlab{}.
\newblock \bibinfo{title}{{NVIDIA DGX GH200}}.
\newblock
  \bibinfo{howpublished}{\url{https://www.nvidia.cn/data-center/dgx-gh200/}}.
\newblock
\urldef\tempurl%
\url{https://www.nvidia.cn/data-center/dgx-gh200/}
\showURL{%
\tempurl}


\bibitem[NVIDIA(2024a)]%
        {megatron-ckpt}
\bibfield{author}{\bibinfo{person}{NVIDIA}.} \bibinfo{year}{2024}\natexlab{a}.
\newblock \bibinfo{title}{The checkpointing of Megatron-LM}.
\newblock
  \bibinfo{howpublished}{\url{https://github.com/NVIDIA/Megatron-LM/blob/main/megatron/core/transformer/transformer_block.py}}.
\newblock
\urldef\tempurl%
\url{https://github.com/NVIDIA/Megatron-LM/blob/main/megatron/core/transformer/transformer_block.py}
\showURL{%
\tempurl}


\bibitem[NVIDIA(2024b)]%
        {megatron}
\bibfield{author}{\bibinfo{person}{NVIDIA}.} \bibinfo{year}{2024}\natexlab{b}.
\newblock \bibinfo{title}{Megatron-LM}.
\newblock
  \bibinfo{howpublished}{\url{https://github.com/NVIDIA/Megatron-LM/tree/main}}.
\newblock
\urldef\tempurl%
\url{https://github.com/NVIDIA/Megatron-LM/tree/main}
\showURL{%
\tempurl}


\bibitem[NVIDIA(2024c)]%
        {megatron-manual}
\bibfield{author}{\bibinfo{person}{NVIDIA}.} \bibinfo{year}{2024}\natexlab{c}.
\newblock \bibinfo{title}{Megatron-LM}.
\newblock \bibinfo{howpublished}{\url{https://github.com/NVIDIA/Megatron-LM}}.
\newblock
\urldef\tempurl%
\url{https://github.com/NVIDIA/Megatron-LM}
\showURL{%
\tempurl}


\bibitem[NVIDIA(2024d)]%
        {b200}
\bibfield{author}{\bibinfo{person}{NVIDIA}.} \bibinfo{year}{2024}\natexlab{d}.
\newblock \bibinfo{title}{{NVIDIA DGX B200}}.
\newblock
  \bibinfo{howpublished}{\url{https://www.nvidia.com/en-us/data-center/dgx-b200/}}.
\newblock
\urldef\tempurl%
\url{https://www.nvidia.com/en-us/data-center/dgx-b200/}
\showURL{%
\tempurl}


\bibitem[NVIDIA(2024e)]%
        {superpod}
\bibfield{author}{\bibinfo{person}{NVIDIA}.} \bibinfo{year}{2024}\natexlab{e}.
\newblock \bibinfo{title}{{NVIDIA's DGX SuperPOD cloud-native supercomputer}}.
\newblock
  \bibinfo{howpublished}{\url{https://www.nvidia.com/en-us/data-center/dgx-superpod-gb200/}}.
\newblock
\urldef\tempurl%
\url{https://www.nvidia.com/en-us/data-center/dgx-superpod-gb200/}
\showURL{%
\tempurl}


\bibitem[Peng et~al\mbox{.}(2020)]%
        {Capuchin-ASPLOS20}
\bibfield{author}{\bibinfo{person}{Xuan Peng}, \bibinfo{person}{Xuanhua Shi},
  \bibinfo{person}{Hulin Dai}, \bibinfo{person}{Hai Jin},
  \bibinfo{person}{Weiliang Ma}, \bibinfo{person}{Qian Xiong},
  \bibinfo{person}{Fan Yang}, {and} \bibinfo{person}{Xuehai Qian}.}
  \bibinfo{year}{2020}\natexlab{}.
\newblock \showarticletitle{{Capuchin: Tensor-based GPU Memory Management for
  Deep Learning}}. In \bibinfo{booktitle}{\emph{Proceedings of the
  International Conference on Architectural Support for Programming Languages
  and Operating Systems}}.
\newblock


\bibitem[PyTorch(2020)]%
        {PyTorch}
\bibfield{author}{\bibinfo{person}{PyTorch}.} \bibinfo{year}{2020}\natexlab{}.
\newblock \bibinfo{title}{PyTorch/Vision}.
\newblock
  \bibinfo{howpublished}{\url{https://github.com/pytorch/vision/tree/master/torchvision}}.
\newblock
\urldef\tempurl%
\url{https://github.com/pytorch/vision/tree/master/torchvision}
\showURL{%
\tempurl}


\bibitem[PyTorch(2024)]%
        {GPU_Util}
\bibfield{author}{\bibinfo{person}{PyTorch}.} \bibinfo{year}{2024}\natexlab{}.
\newblock \bibinfo{title}{{Gpu utilization Kineto}}.
\newblock
  \bibinfo{howpublished}{\url{https://github.com/pytorch/kineto/blob/main/tb_plugin/docs/gpu_utilization.md}}.
\newblock
\urldef\tempurl%
\url{https://github.com/pytorch/kineto/blob/main/tb_plugin/docs/gpu_utilization.md}
\showURL{%
\tempurl}


\bibitem[Radford et~al\mbox{.}(2019)]%
        {gpt2}
\bibfield{author}{\bibinfo{person}{Alec Radford}, \bibinfo{person}{Jeffrey Wu},
  \bibinfo{person}{Rewon Child}, \bibinfo{person}{David Luan},
  \bibinfo{person}{Dario Amodei}, \bibinfo{person}{Ilya Sutskever},
  {et~al\mbox{.}}} \bibinfo{year}{2019}\natexlab{}.
\newblock \showarticletitle{{Language Models are Unsupervised Multitask
  Learners}}.
\newblock \bibinfo{journal}{\emph{OpenAI blog}} (\bibinfo{year}{2019}).
\newblock


\bibitem[Rashidi et~al\mbox{.}({[n.\,d.]})]%
        {ACE-isca2021}
\bibfield{author}{\bibinfo{person}{S. Rashidi}, \bibinfo{person}{M. Denton},
  \bibinfo{person}{S. Sridharan}, \bibinfo{person}{A. Suresh},
  \bibinfo{person}{J. Nie}, {and} \bibinfo{person}{T. Krishna}.}
  \bibinfo{year}{[n.\,d.]}\natexlab{}.
\newblock \showarticletitle{{Enabling Compute-Communication Overlap in
  Distributed Deep Learning Training Platforms}}. In
  \bibinfo{booktitle}{\emph{Proceedings of the 48th Annual International
  Symposium on Computer Architecture}}.
\newblock


\bibitem[Rhu et~al\mbox{.}(2016)]%
        {VDNN-MICRO16}
\bibfield{author}{\bibinfo{person}{Minsoo Rhu}, \bibinfo{person}{Natalia
  Gimelshein}, \bibinfo{person}{Jason Clemons}, \bibinfo{person}{Arslan
  Zulfiqar}, {and} \bibinfo{person}{Stephen~W. Keckler}.}
  \bibinfo{year}{2016}\natexlab{}.
\newblock \showarticletitle{{VDNN: Virtualized Deep Neural Networks for
  Scalable, Memory-Efficient Neural Network Design}}. In
  \bibinfo{booktitle}{\emph{Proceedings of the Annual International Symposium
  on Microarchitecture}}.
\newblock


\bibitem[Shoeybi et~al\mbox{.}(2019)]%
        {Megatron-Arxiv19}
\bibfield{author}{\bibinfo{person}{Mohammad Shoeybi}, \bibinfo{person}{Mostofa
  Patwary}, \bibinfo{person}{Raul Puri}, \bibinfo{person}{Patrick LeGresley},
  \bibinfo{person}{Jared Casper}, {and} \bibinfo{person}{Bryan Catanzaro}.}
  \bibinfo{year}{2019}\natexlab{}.
\newblock \showarticletitle{{Megatron-LM: Training Multi-billion Parameter
  Language Models Using Model Parallelism}}.
\newblock \bibinfo{journal}{\emph{arXiv preprint arXiv:1909.08053}}
  (\bibinfo{year}{2019}).
\newblock


\bibitem[Sun et~al\mbox{.}(2024)]%
        {AdaPipe}
\bibfield{author}{\bibinfo{person}{Zhenbo Sun}, \bibinfo{person}{Huanqi Cao},
  \bibinfo{person}{Yuanwei Wang}, \bibinfo{person}{Guanyu Feng},
  \bibinfo{person}{Shengqi Chen}, \bibinfo{person}{Haojie Wang}, {and}
  \bibinfo{person}{Wenguang Chen}.} \bibinfo{year}{2024}\natexlab{}.
\newblock \showarticletitle{{AdaPipe: Optimizing Pipeline Parallelism with
  Adaptive Recomputation and Partitioning}}. In
  \bibinfo{booktitle}{\emph{Proceedings of the 29th ACM International
  Conference on Architectural Support for Programming Languages and Operating
  Systems, Volume 3}}.
\newblock


\bibitem[Tarnawski et~al\mbox{.}(2020)]%
        {pp-nips20}
\bibfield{author}{\bibinfo{person}{Jakub~M Tarnawski}, \bibinfo{person}{Amar
  Phanishayee}, \bibinfo{person}{Nikhil Devanur}, \bibinfo{person}{Divya
  Mahajan}, {and} \bibinfo{person}{Fanny Nina~Paravecino}.}
  \bibinfo{year}{2020}\natexlab{}.
\newblock \showarticletitle{{Efficient Algorithms for Device Placement of DNN
  Graph Operators}}.
\newblock \bibinfo{journal}{\emph{Advances in Neural Information Processing
  Systems}}  \bibinfo{volume}{33} (\bibinfo{year}{2020}).
\newblock


\bibitem[Vaswani et~al\mbox{.}(2017)]%
        {attention}
\bibfield{author}{\bibinfo{person}{Ashish Vaswani}, \bibinfo{person}{Noam
  Shazeer}, \bibinfo{person}{Niki Parmar}, \bibinfo{person}{Jakob Uszkoreit},
  \bibinfo{person}{Llion Jones}, \bibinfo{person}{Aidan~N Gomez},
  \bibinfo{person}{{\L}ukasz Kaiser}, {and} \bibinfo{person}{Illia
  Polosukhin}.} \bibinfo{year}{2017}\natexlab{}.
\newblock \showarticletitle{{Attention is All You Need}}. In
  \bibinfo{booktitle}{\emph{Proceedings of the advances in neural information
  processing systems}}.
\newblock


\bibitem[Wang et~al\mbox{.}(2018)]%
        {SuperNeurons-PPoPP18}
\bibfield{author}{\bibinfo{person}{Linnan Wang}, \bibinfo{person}{Jinmian Ye},
  \bibinfo{person}{Yiyang Zhao}, \bibinfo{person}{Wei Wu}, \bibinfo{person}{Ang
  Li}, \bibinfo{person}{Shuaiwen~Leon Song}, \bibinfo{person}{Zenglin Xu},
  {and} \bibinfo{person}{Tim Kraska}.} \bibinfo{year}{2018}\natexlab{}.
\newblock \showarticletitle{{SuperNeurons: Dynamic GPU Memory Management for
  Training Deep Neural Networks}}. In \bibinfo{booktitle}{\emph{Proceedings of
  the ACM SIGPLAN Symposium on Principles and Practice of Parallel
  Programming}}.
\newblock


\bibitem[Wang et~al\mbox{.}(2023)]%
        {pipe-3}
\bibfield{author}{\bibinfo{person}{Shibo Wang}, \bibinfo{person}{Jinliang Wei},
  \bibinfo{person}{Amit Sabne}, \bibinfo{person}{Andy Davis},
  \bibinfo{person}{Berkin Ilbeyi}, \bibinfo{person}{Blake Hechtman},
  \bibinfo{person}{Dehao Chen}, \bibinfo{person}{Karthik Srinivasa~Murthy},
  \bibinfo{person}{Marcello Maggioni}, \bibinfo{person}{Qiao Zhang},
  \bibinfo{person}{Sameer Kumar}, \bibinfo{person}{Tongfei Guo},
  \bibinfo{person}{Yuanzhong Xu}, {and} \bibinfo{person}{Zongwei Zhou}.}
  \bibinfo{year}{2023}\natexlab{}.
\newblock \showarticletitle{{Overlap Communication with Dependent Computation
  via Decomposition in Large Deep Learning Models}}. In
  \bibinfo{booktitle}{\emph{Proceedings of the 28th ACM International
  Conference on Architectural Support for Programming Languages and Operating
  Systems}}.
\newblock


\bibitem[WikiText2(2024)]%
        {WikiText2}
\bibfield{author}{\bibinfo{person}{WikiText2}.}
  \bibinfo{year}{2024}\natexlab{}.
\newblock \bibinfo{title}{{WikiText2}}.
\newblock
  \bibinfo{howpublished}{\url{https://paperswithcode.com/dataset/wikitext-2}}.
\newblock
\urldef\tempurl%
\url{https://paperswithcode.com/dataset/wikitext-2}
\showURL{%
\tempurl}


\bibitem[Xiao et~al\mbox{.}(2020)]%
        {AntMan-OSDI20}
\bibfield{author}{\bibinfo{person}{Wencong Xiao}, \bibinfo{person}{Shiru Ren},
  \bibinfo{person}{Yong Li}, \bibinfo{person}{Yang Zhang},
  \bibinfo{person}{Pengyang Hou}, \bibinfo{person}{Zhi Li},
  \bibinfo{person}{Yihui Feng}, \bibinfo{person}{Wei Lin}, {and}
  \bibinfo{person}{Yangqing Jia}.} \bibinfo{year}{2020}\natexlab{}.
\newblock \showarticletitle{{AntMan: Dynamic Scaling on GPU Clusters for Deep
  Learning}}. In \bibinfo{booktitle}{\emph{Proceedings of the 14th USENIX
  Symposium on Operating Systems Design and Implementation}}.
\newblock


\bibitem[Yan et~al\mbox{.}(2021)]%
        {videogpt}
\bibfield{author}{\bibinfo{person}{Wilson Yan}, \bibinfo{person}{Yunzhi Zhang},
  \bibinfo{person}{Pieter Abbeel}, {and} \bibinfo{person}{Aravind Srinivas}.}
  \bibinfo{year}{2021}\natexlab{}.
\newblock \showarticletitle{{Videogpt: Video Generation Using VQ-VAE and
  Transformers}}.
\newblock \bibinfo{journal}{\emph{arXiv preprint arXiv:2104.10157}}
  (\bibinfo{year}{2021}).
\newblock


\bibitem[Yang et~al\mbox{.}(2021)]%
        {yang2021auto}
\bibfield{author}{\bibinfo{person}{Siling Yang}, \bibinfo{person}{Weijian
  Chen}, \bibinfo{person}{Xuechen Zhang}, \bibinfo{person}{Shuibing He},
  \bibinfo{person}{Yanlong Yin}, {and} \bibinfo{person}{Xian-He Sun}.}
  \bibinfo{year}{2021}\natexlab{}.
\newblock \showarticletitle{{AUTO-PRUNE: Automated DNN Pruning and Mapping for
  ReRAM-based Accelerator}}. In \bibinfo{booktitle}{\emph{Proceedings of the
  ACM International Conference on Supercomputing}}.
\newblock


\bibitem[Zhang et~al\mbox{.}(2022)]%
        {opt-arxiv22}
\bibfield{author}{\bibinfo{person}{Susan Zhang}, \bibinfo{person}{Stephen
  Roller}, \bibinfo{person}{Naman Goyal}, \bibinfo{person}{Mikel Artetxe},
  \bibinfo{person}{Moya Chen}, \bibinfo{person}{Shuohui Chen},
  \bibinfo{person}{Christopher Dewan}, \bibinfo{person}{Mona Diab},
  \bibinfo{person}{Xian Li}, \bibinfo{person}{Xi~Victoria Lin},
  {et~al\mbox{.}}} \bibinfo{year}{2022}\natexlab{}.
\newblock \showarticletitle{{Opt: Open Pre-trained Transformer Language
  Models}}.
\newblock \bibinfo{journal}{\emph{arXiv preprint arXiv:2205.01068}}
  (\bibinfo{year}{2022}).
\newblock


\bibitem[Zhou et~al\mbox{.}(2023)]%
        {MPress-HPCA23}
\bibfield{author}{\bibinfo{person}{Quan Zhou}, \bibinfo{person}{Haiquan Wang},
  \bibinfo{person}{Xiaoyan Yu}, \bibinfo{person}{Cheng Li},
  \bibinfo{person}{Youhui Bai}, \bibinfo{person}{Feng Yan}, {and}
  \bibinfo{person}{Yinlong Xu}.} \bibinfo{year}{2023}\natexlab{}.
\newblock \showarticletitle{{MPress: Democratizing Billion-Scale Model Training
  on Multi-GPU Servers via Memory-Saving Inter-Operator Parallelism}}. In
  \bibinfo{booktitle}{\emph{Proceedings of the 2023 IEEE International
  Symposium on High-Performance Computer Architecture}}.
\newblock


\bibitem[Zong et~al\mbox{.}(2023)]%
        {STR}
\bibfield{author}{\bibinfo{person}{Zan Zong}, \bibinfo{person}{Li Lin},
  \bibinfo{person}{Leilei Lin}, \bibinfo{person}{Lijie Wen}, {and}
  \bibinfo{person}{Yu Sun}.} \bibinfo{year}{2023}\natexlab{}.
\newblock \showarticletitle{{STR: Hybrid Tensor Re-Generation to Break Memory
  Wall for DNN Training}}.
\newblock \bibinfo{journal}{\emph{IEEE Transactions on Parallel and Distributed
  Systems}} (\bibinfo{year}{2023}).
\newblock


\end{thebibliography}
	
	\bibliographystyle{ACM-Reference-Format}

	\appendix
	
\end{document}